%% file: 08_inverse_design_for_dsa.tex
\documentclass[12pt]{spieman}  
\usepackage{amsmath,amsfonts,amssymb}
\usepackage{graphicx}
\usepackage{setspace}
\usepackage{tocloft}
\input{packages.tex}

\input{macros.tex}

\title{A non-parametric shape optimization approach for solving inverse problems in directed self-assembly of block copolymers}

\author[a,*]{Daniil Bochkov}
\author[a,b]{Frederic Gibou}
\affil[a]{Department of Mechanical Engineering, University of California, Santa Barbara, Santa Barbara, CA, USA, 93106}
\affil[b]{Department of Computer Science, University of California, Santa Barbara, Santa Barbara, CA, USA, 93106}

\cftpagenumbersoff{figure}
\cftpagenumbersoff{table} 
\begin{document} 
\maketitle

\begin{abstract}
In this work we consider the inverse problem of finding guiding pattern shapes that result in desired self-assembly morphologies of block copolymer melts. Specifically, we model polymer self-assembly using Self-Consistent Field Theory and derive in a non-parametric setting the sensitivity of the misfit between desired and actual morphologies to arbitrary perturbations in the guiding pattern shape. The obtained sensitivities are used for optimization of the confining pattern shapes such that the misfit between desired and actual morphologies is minimized. The efficiency and robustness of the proposed algorithm is demonstrated on a number of examples related to templating Vertical Interconnect Accesses.
\end{abstract}


{\noindent \footnotesize\textbf{*}Daniil Bochkov,  \linkable{bochkov.ds@gmail.com} }

\begin{spacing}{2}   

\section{Introduction}

The self-assembling characteristics of block copolymer melts have the potential to further enhance the nanofabrication techniques of electronic components\cite{jeong2013directed,hu2014directed}. Typically, the polymer self-assembly needs to be guided using external physical and chemical stimuli to achieve desired template morphologies\cite{darling2007directing}. It is relatively straightforward to predict the self-assembling polymer structure in a given guiding design using either the Ohta-Kawasaki phase-field model\cite{ohta1986equilibrium} or more accurate Self-Consistent Field Theory\cite{fredrickson2006equilibrium}. However, the inverse problem to that, i.e., finding guiding desing that results in a desired morphologies, could be significantly more challenging due to the highly nonlinear nature of the mathematical theories describing the polymer self-assembly. In certain applications, this problem can be easily solved by a simple trial and error approach given the low dimensionality of the control variable, for example, the distance between guiding features in density multiplication techniques\cite{ruiz2008density,bita2008graphoepitaxy,liu2013chemical}. However, in other applications the control variable can be of very high finite dimensionality, for example, positions of guide posts for generating complex structures \cite{hannon2013inverse, hannon2014optimizing}, or even infinitely dimensional, for example, the shape of confining mask for templating Vertical Interconnect Accesses (VIAs)\cite{tiron2013potential, ouaknin2016shape, ouaknin2018level}. Advanced solution techniques are required in such situations.

In the context of obtaining optimized shapes of confining mask for templating of VIAs, this challenging inverse problem has been previously approached, for example, by using SCFT and parameterizing the confinement geometry with several degrees of freedom\cite{latypov2013computational, shim2016mask}. The derivatives of the cost functional with respect to such degrees of freedom, which are needed for an efficient optimization, were approximated either using a brute-force finite difference approach or a linearization of SCFT equations. In another work \cite{ouaknin2016shape, ouaknin2018level}, a non-parametric description of the confining mask was employed leveraging the Level-Set Method\cite{sethian1999level,osher2003level}, a powerful methodology for implicit representation and manipulation of arbitrary geometries. However, the optimization problem was solved only approximately using intuitive physical arguments. In this work, we propose an approach that combines the strengths of previous works. Specifically, we model the polymer self-assembly using SCFT and derive in a non-parametric setting the exact sensitivities of the cost functional with respect to arbitrary deformation in the shape of confining mask. Using this information in conjunction with the machinery of the Level-Set Method, we efficiently optimize the shapes of confining masks with respect to the misfit between actual and desired morphologies.

The rest of this manuscript is organized in the following way. Section \ref{sec:scft} briefly introduces governing equations of the Self-Consistent Field Theory that describes self-assembly of block copolymer melts. In Section \ref{sec:dsa} we derive non-parametric sensitivities of the self-assembled morphology to the shape of guide pattern. In Section \ref{sec:numerical} we discuss numerical aspects of this work. Section \ref{sec:results} contains examples of applying the proposed methodology in different situations. Finally, Section \ref{sec:conclusion} concludes the manuscript.

\section{Mathematical model of block copolymer self-assembly}\label{sec:scft}

In this work we employ Self-Consistent Field Theory\cite{fredrickson2006equilibrium} for modeling self-assembly of block copolymer melts. Let us consider an AB-type diblock copolymer melt confined in a guide pattern of shape $\Omega$ with boundary $\Gamma$. Using a common notation, we denote the total number of statistical segments in each polymer chain as $N$, the fraction of segments belonging to chemical species A as $f$, and the Flory-Higgins parameter characterizing the interaction strength between species A and B as $\chi$. As shown in \cite{bochkov2021equilibrium}, the total energy per polymer chain of such a system can be written as
\begin{myalign*}
\ham = \frac{1}{V} 
\myint_{\Omega} \left( \frac{\mu_-}{\chi N} - \mu_+ \right) \diff{\vect{x}}
-\log\of{\Q} + \sigma \myint_{\Gamma} \left( \rho_A \gamma_A + \rho_B \gamma_B \right) \diff{\Gamma}
\end{myalign*}
where $\mu_+$ and $\mu_-$ denote the fluctuating pressure and exchange chemical fields, $V$ is the volume of the guide pattern, $\Q$ is the single-chain partition function, $\sigma = \frac{N v_0}{R_g} \frac{1}{k_B T}$ is a constant factor, $\gamma_A$ and $\gamma_B$ are surface energies between walls of the guide patterns and chemical species A and B, $\rho_A$ and $\rho_B$ are the density fields of block A and B. The signle-chain partition function $\Q$ and density fields $\rho_A$, $\rho_B$ can be calculated using the so-called forward and backward chain propagators $\qf=\qf\of{s,\vect{x}}$ and $\qb=\qb\of{s,\vect{x}}$
\begin{myalign*}
\Q &= \frac{1}{V} \myint_\Omega \qf\of{1,\vect{x}} \diff{\vect{x}},
\\
\rho_A &= \frac{1}{Q} \myint_{0}^{f} \qf\of{s,\vect{x}} \qb\of{s,\vect{x}}, \diff{s}
\\
\rho_B &= \frac{1}{Q} \myint_{f}^{1} \qf\of{s,\vect{x}} \qb\of{s,\vect{x}}, \diff{s}
\end{myalign*}
which, in their turn satisfy the following modified diffusion equations
\begin{myalign}
\label{eq:diff:fwd}
\left\{
\begin{aligned}
\dds{\qf} + \mu\of{s} \qf &= \lap \qf, 
&\vect{r} &\in \Omega, & 0 &< s \leq 1, \\
\ddn{\qf} + \sigma \left( \gamma\of{s} - \gamma_{\text{c}} \right) \qf&= 0, 
&\vect{r} &\in\Gamma, & 0 &< s \leq 1, \\
\qf &= 1, 
&\vect{r} &\in \Omega, & s &= 0
\end{aligned}
\right.
\end{myalign}
and
\begin{myalign}
\label{eq:diff:bwd}
\left\{
\begin{aligned}
-\dds{\qb} + \mu\of{s} \qb &= \lap \qb, 
&\vect{r} &\in \Omega, & 0 &\leq s < 1, \\
\ddn{\qb} + \sigma \left( \gamma\of{s} - \gamma_{\text{c}} \right)\qb &= 0, 
&\vect{r} &\in\Gamma, & 0 &\leq s < 1, \\
\qb &= 1, 
&\vect{r} &\in \Omega, & s &= 1,
\end{aligned}
\right.
\end{myalign}
where $\gamma_{\text{c}} = \rho_A \gamma_A + \rho_B \gamma_B$,
\begin{myalign*}
\mu\of{s} =
\begin{cases}
\mu_+ - \mu_-, & s < f \\
\mu_+ + \mu_-, & s > f
\end{cases}
, \quad \text{and} \quad
\gamma\of{s} =
\begin{cases}
\gamma_A, & s < f \\
\gamma_B, & s > f
\end{cases}
\end{myalign*}
Note that the terms containing $\gamma_{\text{c}}$ in the Robin boundary conditions above ensures the consistency with the incompressibility of the melt as discussed in \cite{bochkov2021equilibrium}.

The stable polymer morphologies correspond to saddle points of the energy functional $\ham$ and satisfy 
\begin{myalign}
\label{eq:pressure}
\fd{\ham}{\mu_+} &= \rho_A + \rho_B - 1 = 0,\\
\label{eq:exchange}
\fd{\ham}{\mu_-} &= \frac{2\mu_-}{\chi N} -\rho_A + \rho_B = 0.
\end{myalign}
Given the nonlinear character of the theory such solution are typically obtained iteratively using steepest descent/ascent methods. Note that in the formulation used in this work density fields $\rho_A$ and $\rho_B$ are calculated through $\qf$ and $\qb$, while boundary conditions for $\qf$ and $\qb$ themselves depend on $\rho_A$ and $\rho_B$. An iterative solution of such a non-linearity is discussed in \cite{}.

\section{Sensitivity of polymer morphology to confining mask's geometry}\label{sec:dsa}
The SCFT provides the means for obtaining BCP density fields for any confining geometry, that is, for solving the forward problem. In order to solve the inverse problem, that is, find the confining geometry that results in a specified density profile, we define a cost functional that measures the deviation between an actual density field and a desired one and analytically derive its sensitivity to the confinement shape. 
\begin{figure}[!h]
\centering
\includegraphics[width=.3\textwidth]{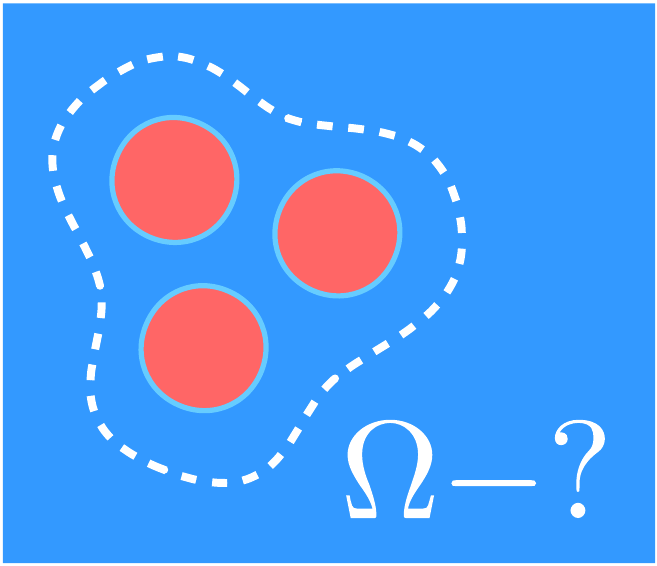}
\caption{Problem geometry and notation used in section \ref{sec:dsa}. The desired density field configuration is represented by the three disks.}
\label{fig:dsa:problem-statement}
\end{figure}
Specifically, noting that the density fields $\rho_A$ and $\rho_B$ are directly related to the exchange field $\mum$ through the SCFT equation \eqref{eq:exchange}, we define the cost functional as:
\begin{myalign}\label{eq:cost}
\cost = \myint_{\Omega} \frac{1}{\XN{AB}} \left( \mum - \mu_t \right)^2 \diff{\vecr} + \alpha \myint_{\Gamma} \diff{\Gamma},
\end{myalign}
where $\Omega$ and $\Gamma$ denotes the confining geometry and its boundary, $\mu_t$ is a desired density field configuration (see figure \ref{fig:dsa:problem-statement}), and $\alpha$ is the curvature penalization parameter. The factor $\frac{1}{\XN{AB}}$ is introduced solely for the sake of convenience in further calculations. Since the efficiency and viability of the directed self-assembly is based on the ability to produce ordered nanostructures of characteristic dimensions smaller than those of guiding masks, it is important to have the means for controlling the maximum allowable curvature of the confining masks. To do so, we introduce a term penalizing for the total perimeter of the shape into the cost functional. However, we note that more intricate strategies can potentially be employed as well. 

To derive the sensitivity of the cost functional \ref{eq:cost} with respect to deformations in confining shape $\Omega$ it is convenient to consider a situation in which the confining boundary $\Gamma$ is evolving with some normal velocity $\vn = \vn\of{\tau, \vecr}$ in fictitious time $\tau$ and compute the derivative of the cost functional with respect to $\tau$. In doing so, one must take into account the fact that changes in geometry of confining mask $\Omega$ leads to changes in chain propagators, $\qf$ and $\qb$, and chemical fields, $\mu_+$ and $\mu_-$, according to diffusion equations \eqref{eq:diff:fwd}-\eqref{eq:diff:bwd} and equilibrium conditions \eqref{eq:pressure}-\eqref{eq:exchange}. Treating these equations as constraints, we define a Lagrangian functional according to
\begin{mymultline*}
\lag = 
\myint_\Omega 
\frac{\left( \mu_- - \mu_t \right)^2}{\XN{AB}} 
\diff{\vect{r}}
+ \alpha \myint_{\Gamma} \diff{\Gamma}
+
\myint_\Omega
\lambda_+
\left[
V \myint_0^1 q q_c \diff{s} - \myint_\Omega \myint_0^1 q q_c \diff{s} \diff{\vect{r}^\prime} 
\right]
\diff{\vect{r}}
\\
+
\myint_\Omega
\lambda_-
\left[
\frac{2}{\chi N} \mu_- \myint_\Omega \myint_0^1 q q_c \diff{s} \diff{\vect{r}^\prime} + V \myint_0^1 \sgn\of{s-f} q q_c \diff{s} 
\right]
\diff{\vect{r}}
\\
+
\myint_{\Omega}
\left(
q(1) \lambda_c(1) - \lambda_c(0)
\right)
\diff{\vect{r}} 
+
\myint_{\Omega}
\myint_{0}^1
\left(
-q \partial_s \lambda_c + \mu q \lambda_c + \nabla q \cdot \nabla \lambda_c
\right)
\diff{s}
\diff{\vect{r}}
\\
+
\myint_{\Omega}
\left(
q_c(0) \lambda(0) - \lambda(1)
\right)
\diff{\vect{r}} 
+
\myint_{\Omega}
\myint_{0}^1
\left(
q_c \partial_s \lambda + \mu q_c \lambda + \nabla q_c \cdot \nabla \lambda
\right)
\diff{s}
\diff{\vect{r}}
\\
+  
\myint_{\Gamma}
\myint_{0}^1
\sigma (\gamma\of{s} - \gamma_{\text{c}}) \qf \lb
\diff{s}
\diff{\vect{r}}
+  
\myint_{\Gamma}
\myint_{0}^1
\sigma (\gamma\of{s} - \gamma_{\text{c}}) \qb \lf
\diff{s}
\diff{\vect{r}},
\end{mymultline*} 
where $\lf$, $\lb$, $\lm$, and $\lp$ are the Lagrange multipliers. Note that equilibrium conditions \eqref{eq:pressure}-\eqref{eq:exchange} are incorporated after explicitly expressing density fields through $\qf$ and $\qb$, and diffusion equations \eqref{eq:diff:fwd}-\eqref{eq:diff:bwd} are incorporated using their weak forms. The full derivative of the cost functional with respect to $\tau$ is equal to the partial derivative of the above Lagrangian provided the following optimality conditions are satisfied:
\begin{alignat*}{2}
\delta_{\qf} \lag = \myint_{0}^{1}\myint_\Omega \fd{\lag}{\qf} \delta \qf \diff{\vect{x}} \diff{s} &= 0\, \forall\,\delta\qf, 
\quad& 
\delta_{\lf} \lag = \myint_{0}^{1}\myint_\Omega \fd{\lag}{\lf} \delta \lf \diff{\vect{x}} \diff{s} &= 0\, \forall\,\delta\lf,  
\\
\delta_{\qb} \lag = \myint_{0}^{1}\myint_\Omega \fd{\lag}{\qb} \delta \qb \diff{\vect{x}} \diff{s} &= 0\, \forall\,\delta\qb, 
\quad& 
\delta_{\lb} \lag = \myint_{0}^{1}\myint_\Omega \fd{\lag}{\lb} \delta \lb \diff{\vect{x}} \diff{s} &= 0\, \forall\,\delta\lb,  
\\
\delta_{\mup} \lag = \myint_\Omega \fd{\lag}{\mup} \delta \mup \diff{\vect{x}} &= 0\, \forall\,\delta\mup, 
\quad& 
\delta_{\lp} \lag = \myint_\Omega \fd{\lag}{\lp} \delta \lp \diff{\vect{x}} &= 0\, \forall\,\delta\lp,   
\\
\delta_{\mum} \lag = \myint_\Omega \fd{\lag}{\mum} \delta \mum \diff{\vect{x}} &= 0\, \forall\,\delta\mum, 
\quad& 
\delta_{\lm} \lag = \myint_\Omega \fd{\lag}{\lm} \delta \lm \diff{\vect{x}} &= 0\, \forall\,\delta\lm.
\end{alignat*}
It is trivial to show that the conditions from the right column simply recover equations \eqref{eq:diff:fwd}, \eqref{eq:diff:bwd}, \eqref{eq:pressure}, and \eqref{eq:exchange}, respectively. The conditions from the left column provide the equations for the Lagrange multipliers. Taking the corresponding variations of the Lagrangian and equating them to zero, one obtains the following integral equations:
\begin{mymultline*}
\delta_{\qf} \lag = 
\myint_{\Omega}
\myint_{0}^1
\left(
-\delta q \partial_s \lb + \mu \delta q \lambda_c + \mu_\lambda \delta \qf \qb + \nabla \delta\qf \cdot \nabla \lb
\right)
\diff{s}
\diff{\vect{r}}
\\
+  
\myint_{\Omega}
\delta q(1) \lambda_c(1)
\diff{\vect{r}} 
+
\myint_{\Gamma}
\myint_{0}^1
\sigma (\gamma\of{s} - \gamma_{\text{c}}) \delta \qf \lb
\diff{s}
\diff{\vect{r}}
= 0, \quad \forall\, \delta \qf,
\end{mymultline*} 
\begin{mymultline*}
\delta_{\qb} \lag = 
\myint_{\Omega}
\myint_{0}^1
\left(
\delta \qb \partial_s \lf + \mu \delta \qb \lf + \mu_\lambda \qf \delta \qb + \nabla \delta\qb \cdot \nabla \lf
\right)
\diff{s}
\diff{\vect{r}}
\\
+  
\myint_{\Omega}
\delta \qb(0) \lf(0)
\diff{\vect{r}} 
+
\myint_{\Gamma}
\myint_{0}^1
\sigma (\gamma\of{s} - \gamma_{\text{c}}) \lf \delta \qb
\diff{s}
\diff{\vect{r}} = 0, \quad \forall\, \delta \qb,
\end{mymultline*} 
\begin{mymultline*}
\delta_{\mum} \lag = 
\myint_\Omega 
2\frac{\mu_- - \mu_t}{\XN{AB}} \delta \mu_t
\diff{\vect{r}}
+
\myint_\Omega
\frac{2}{\chi N} \lambda_- \myint_\Omega \myint_0^1 q q_c \diff{s} \diff{\vect{r}^\prime}
\delta \mu_- 
\diff{\vect{r}}
\\
+
\myint_{\Omega}
\myint_{0}^1 \sgn\of{s-f} \left( \qf \lb + \qb \lf \right) \delta \mum 
\diff{s}
\diff{\vect{r}}
=0, \quad \forall\, \delta \mum,
\end{mymultline*} 
\begin{myalign*}
\delta_{\mup} \lag = 
\myint_{\Omega}
\myint_{0}^1 \left( \qf \lb + \qb \lf \right) 
\diff{s}
\delta \mup
\diff{\vect{r}}
=0, \quad \forall\, \delta \mup,
\end{myalign*}
where
\begin{myalign*}
\mu_\lambda = \lambda_+ - \left\langle \lambda_+ \right\rangle  + \sgn\of{s-f} \lambda_- + \left\langle \frac{2\mu_- \lambda_-}{\chi N} \right\rangle.
\end{myalign*}
Using the divergence theorem, it can be shown that the first two of the above equations are weak forms of the following diffusion equations for $\lf$ and $\lb$:
\begin{myalign}
\label{eq:adjoint:diff:fwd}
\left\{
\begin{aligned}
\partial_s \lf + \mu\of{s} \lf + \mu_\lf\of{s} q &= \lap \lf, \\
 \ddn{\lf} + \sigma \left( \gamma\of{s} - \gamma_{\text{c}} \right) \lf &= 0, \\
\lf\of{0,\vecr} &= 0,
\end{aligned}
\right.
\end{myalign}
and
\begin{myalign}
\label{eq:adjoint:diff:bwd}
\left\{
\begin{aligned}
-\partial_s \lb + w \lb + w_\lambda \qb &=  \lap \lb, \\
 \ddn{\lb} + \sigma \left( \gamma\of{s} - \gamma_{\text{c}} \right) \lb &= 0, \\
\lb\of{1,\vecr} &= 0.
\end{aligned}
\right.
\end{myalign}
The latter two integral equations are weak formulations of
\begin{myalign}
\label{eq:adjoint:pressure}
\rho^\lambda_A\of{\vecr} + \rho^\lambda_B\of{\vecr} &= 0, 
\\
\label{eq:adjoint:exchange}
\rho^\lambda_A\of{\vecr} - \rho^\lambda_B\of{\vecr} &= 2\frac{\lambda_-\of{\vecr} + \mu_-\of{\vecr} - \mu_t\of{\vecr}}{\XN{AB}},
\end{myalign}
where $\rho^\lambda_A$ and $\rho^\lambda_B$ are defined as
\begin{myalign*}
\rho^\lambda_A\of{\vecr} = \frac{1}{\Q} \myint_0^f \left( q \lambda_c + q_c \lambda \right) \diff{s},
\\
\rho^\lambda_B\of{\vecr} = \frac{1}{\Q} \myint_f^1 \left( q \lambda_c + q_c \lambda \right) \diff{s}.
\end{myalign*}
As one can see, the Lagrange multipliers $\lf$, $\lb$, $\lm$, and $\lp$ satisfy a system of equations \eqref{eq:adjoint:diff:fwd},
\eqref{eq:adjoint:diff:bwd}, \eqref{eq:adjoint:pressure}, \eqref{eq:adjoint:exchange} that has a n analogous structure as the original system of the SCFT equations \eqref{eq:diff:fwd},
\eqref{eq:diff:bwd}, \eqref{eq:pressure}, \eqref{eq:exchange}. Thus, we employ the same steepest descent/ascent strategy to solve for $\lf$, $\lb$, $\lm$, and $\lp$.

Finally, taking the partial derivative of $\lag$ with respect to $\tau$ produces (see, e.g., \cite{} for differentiation formulas for integral quantities):
\begin{mymultline}\label{eq:dsa:derivative}
\deriv{\cost}{\tau} =		
\pd{\lag}{\tau} = 
\myint_\Gamma
\Bigg\lbrace
\frac{\left( \mu_-(\vect{r}) - \mu_t(\vect{r}) \right)^2}{\chi N} 
- \frac{\lambda(1) + \lambda_c(0)}{2Q}
+ \frac{2\mu_- \lambda_-}{\chi N} - \left\langle \frac{2\mu_- \lambda_-}{\chi N} \right\rangle 
\\
- \lambda_+ + \left\langle \lambda_+ \right\rangle 
+ \alpha \kappa
\Bigg\rbrace
\vn
\diff{\Gamma},	
\end{mymultline}
where $\kappa$ is the mean curvature of the confining geometry.
The derived expression can be used for finding the confinement shapes that results in the closet match between the desired configuration $\mu_t$ and the actual one $\mum$ by iteratively evolving the shapes under the velocity field:
\begin{myalign*}
\vn = -\Bigg\lbrace
\frac{\left( \mu_-(\vect{r}) - \mu_t(\vect{r}) \right)^2}{\chi N} 
- \frac{\lambda(1) + \lambda_c(0)}{2Q}
+ \frac{2\mu_- \lambda_-}{\chi N} - \left\langle \frac{2\mu_- \lambda_-}{\chi N} \right\rangle 
- \lambda_+ + \left\langle \lambda_+ \right\rangle 
+ \alpha \kappa
\Bigg\rbrace.
\end{myalign*}

\section{Numerical aspects}\label{sec:numerical}
Up until this point, the presented approach does not rely on any specific numerical methods. Thus, it could be implemented in any numerical framework advanced enough to solve diffusion equations with Robin boundary conditions in irregular domains and handle geometry evolution.
In this work, we use a combination of the Level-Set Method\cite{sethian1999level,osher2003level}, adaptive Cartesian grids and sharp-interface finite-volume methods for solving PDE along the lines of \cite{ouaknin2016self}. Diffusion equations \eqref{eq:diff:fwd}-\eqref{eq:diff:bwd} are solved with a finite-volume method presented in \cite{bochkov2019solving}. Manipulation of problem geometry is performed by using a semi-implicit advection scheme as  described in \cite{smereka2003semi,min2007second}.

\section{Results}\label{sec:results}
In this section, we present a number of numerical examples to validate the proposed framework and demonstrate its capabilities.

In all the numerical examples of this section, we consider a cylinder-forming diblock copolymer described by parameters $f=0.3$ and $\XN{AB} = 30$. All target patterns consist of cylindrical domains and the target field $\mu_t$ of each cylindrical domain is taken as:
\begin{myalign*}
\mu_t = \hf \XN{AB} \textrm{sgn}\of{\phi} \left( \exp\of{- \abs{\phi} \sqrt{\XN{AB}} } - 1\right),
\end{myalign*}
where
\begin{myalign*}
\phi\of{x,y} = \sqrt{\left(x-x_c\right)^2 + \left(y-y_c\right)^2} - r_0,
\end{myalign*}
and $r_0$ and $(x_c, y_c)$ are the target domain's radius and center, respectively.

\subsection{Validation test}\label{sec:results:dsa:test}
\begin{figure}[!h]
\centering
\includegraphics[width=.15\textwidth]{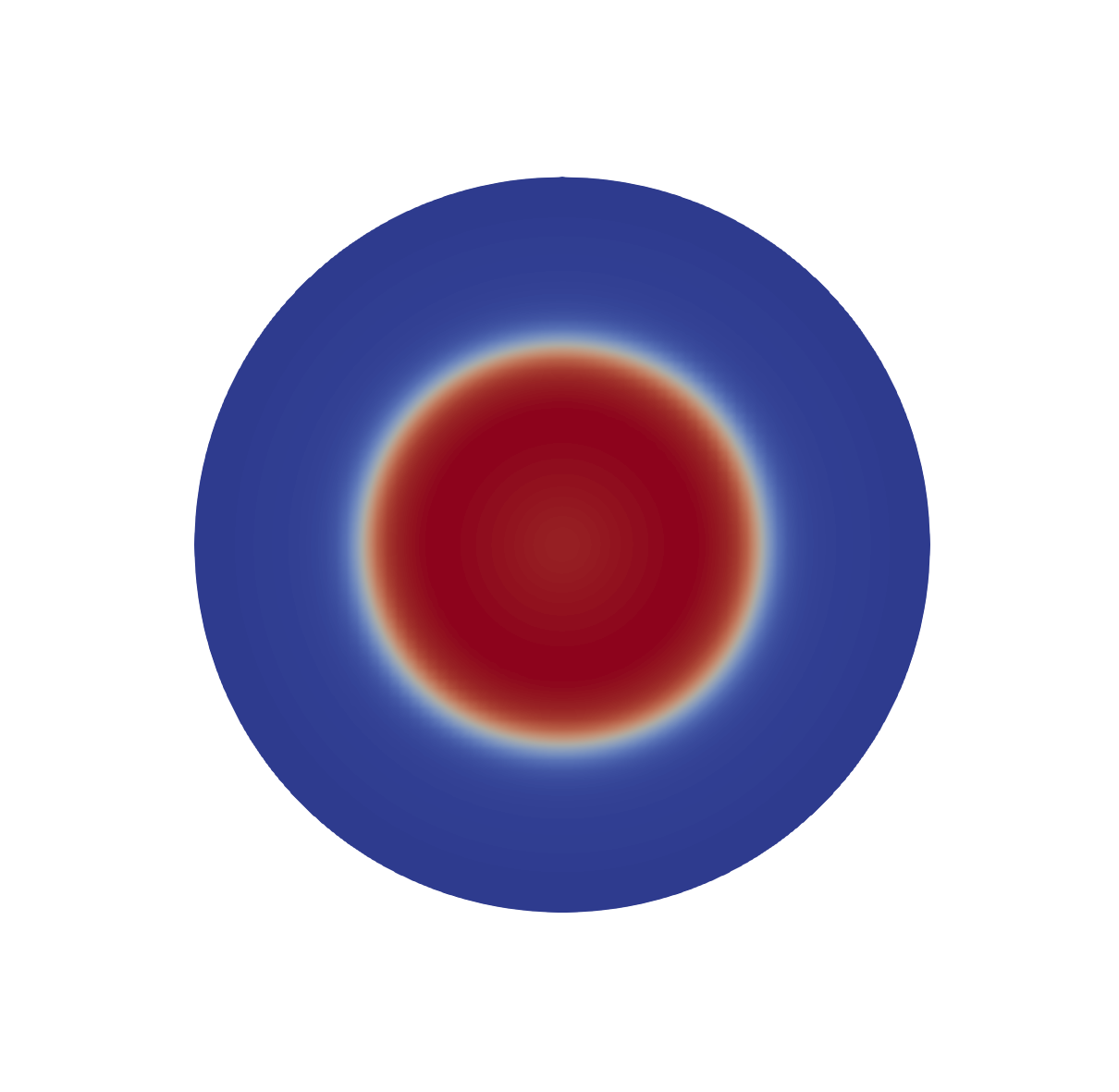}
\includegraphics[width=.15\textwidth]{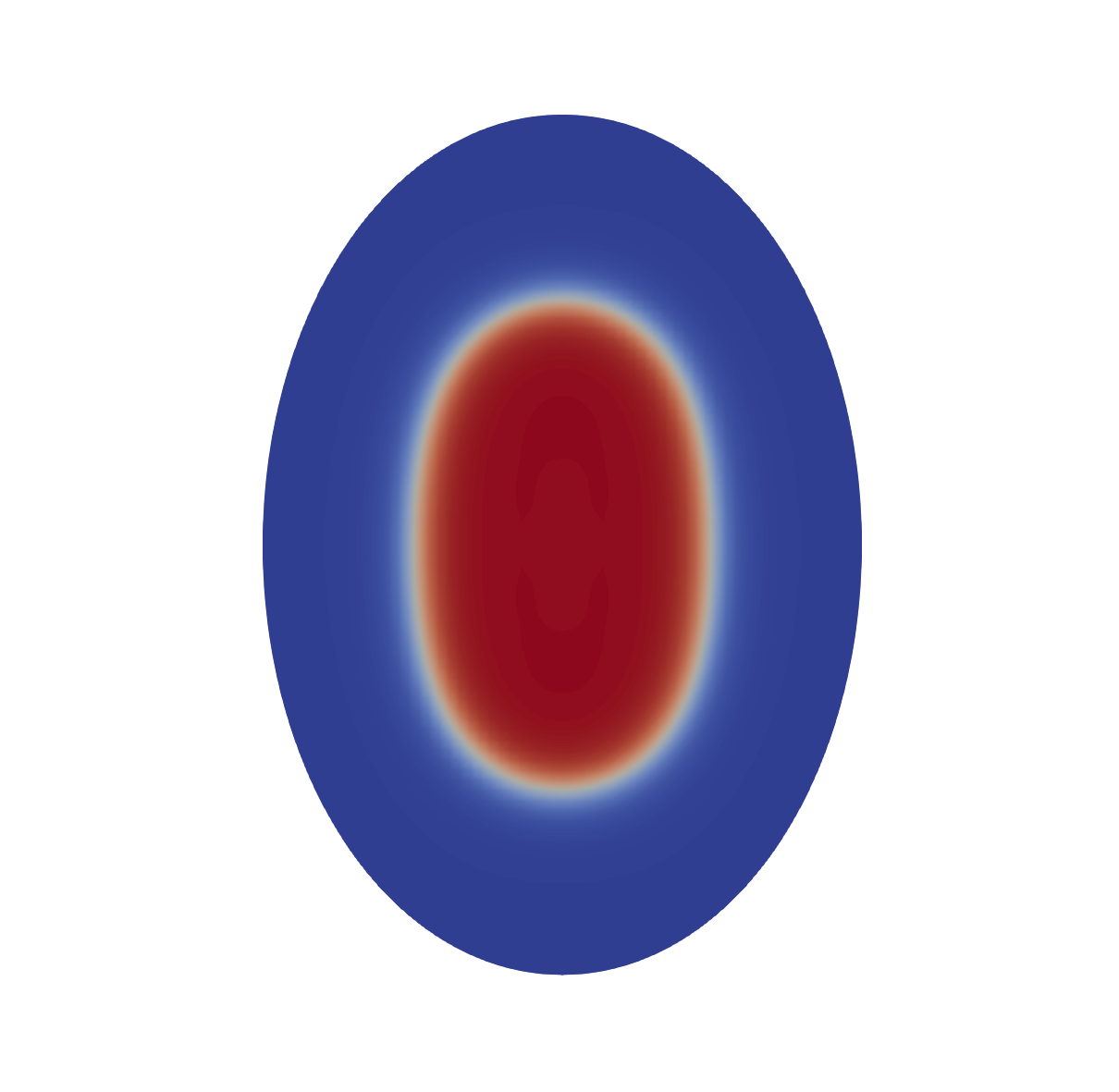}
\includegraphics[width=.15\textwidth]{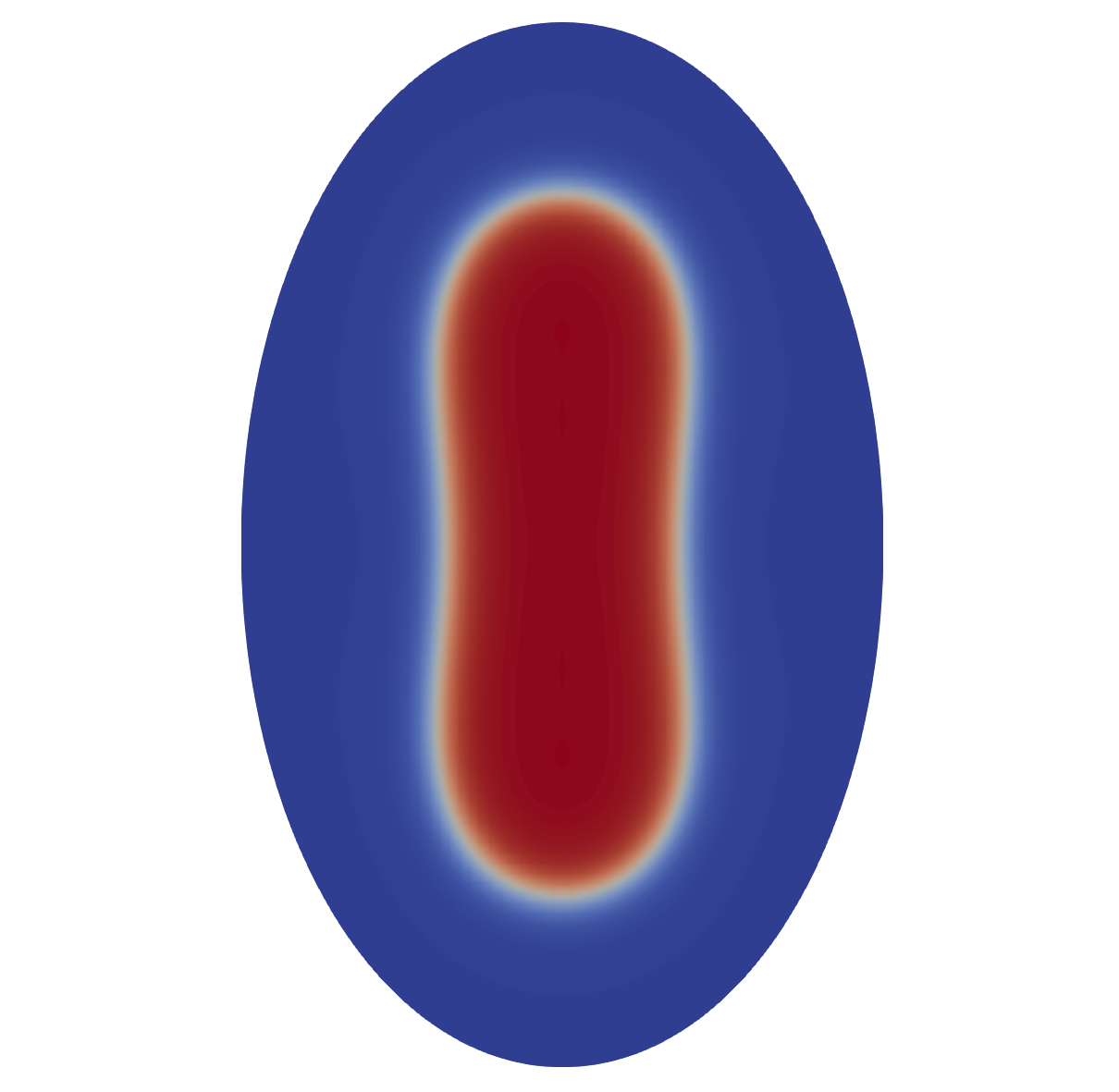}
\includegraphics[width=.15\textwidth]{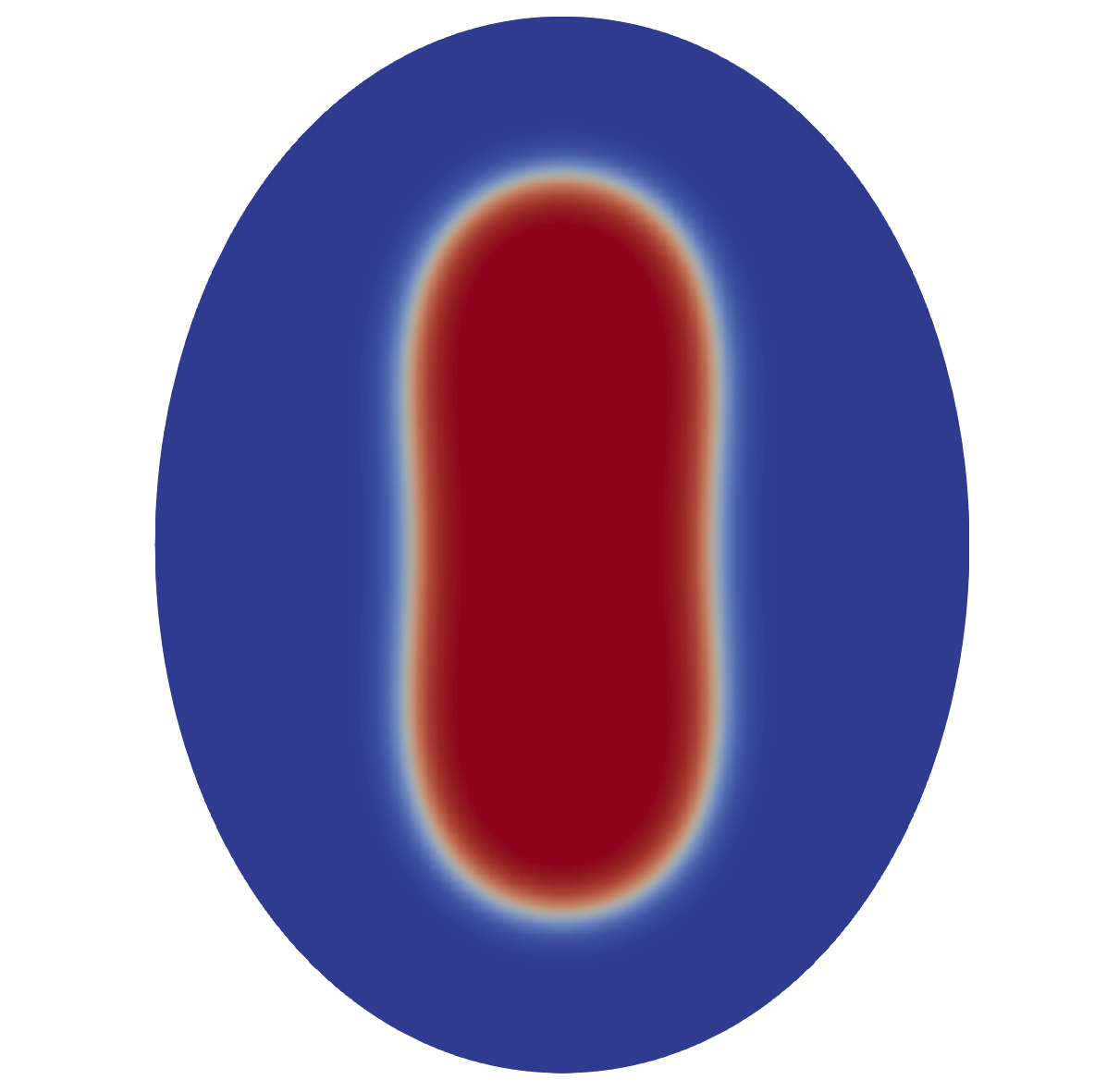}
\includegraphics[width=.15\textwidth]{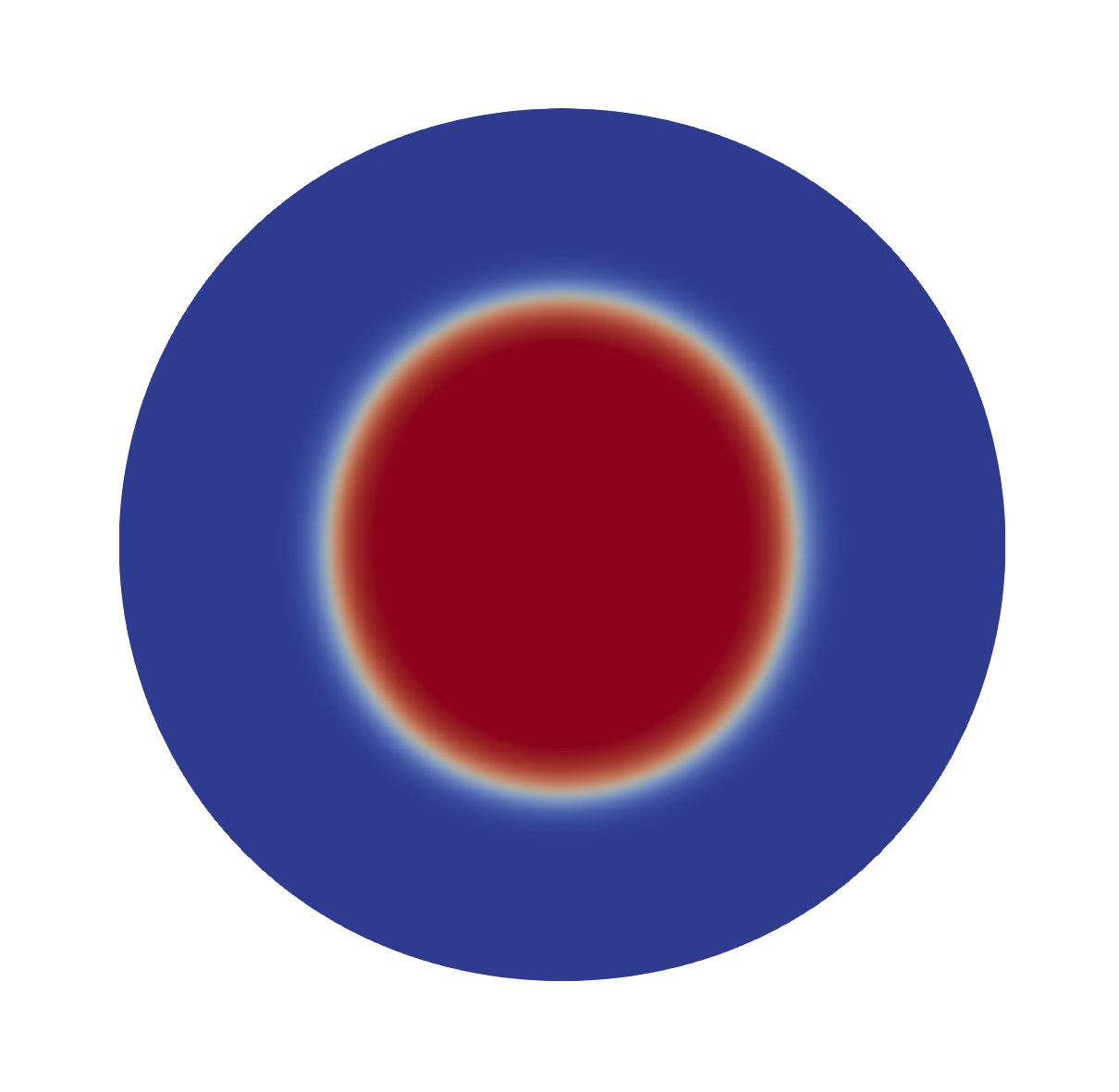}
\includegraphics[width=.15\textwidth]{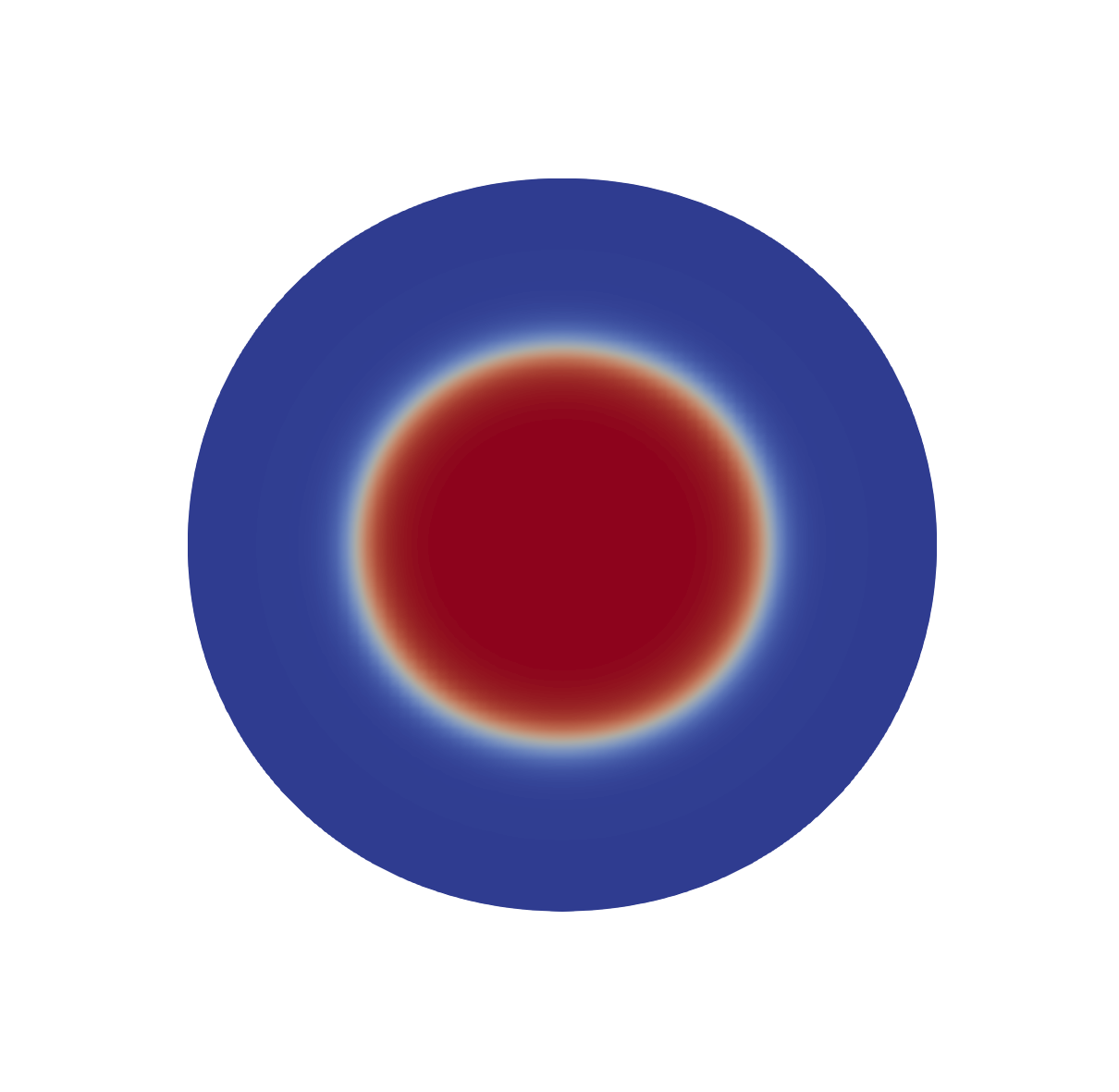}
\caption{Visualization of imposed motion in example \ref{sec:results:dsa:test}.}
\label{fig:results:dsa:test:visual}
\end{figure}

\begin{figure}[!h]
\centering
\includegraphics[width=.99\textwidth]{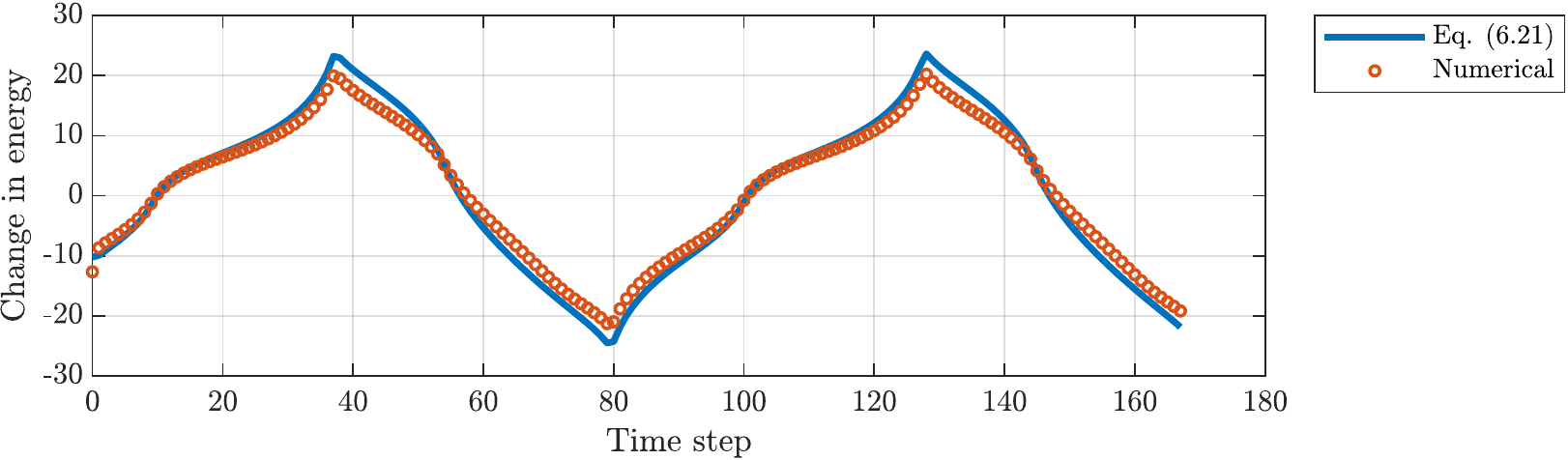}
\caption{Comparison between changes in energy computed using \eqref{eq:dsa:derivative} and numerically.}
\label{fig:results:dsa:test:compare}
\end{figure}

Before presenting some practical applications of the proposed algorithm for the inverse design problem in DSA, we first validate it on a synthetic test. Specifically, we consider a circular confining domain of radius $3.5 R_g$, a circular target domain of radius $1.0 R_g$, and impose deformation velocities according to 
\begin{myalign*}
\begin{aligned}
v_x &=  x \cos\of{5\tau}, \\
v_y &= -y \cos\of{5\tau}.
\end{aligned}
\end{myalign*}
Figure \ref{fig:results:dsa:test:visual} illustrate the self-assembly under an imposed deformation of the surface. Figure \ref{fig:results:dsa:test:compare} shows the comparison between the change in the cost functional calculated directly and as predicted by the analytical expression \eqref{eq:dsa:derivative}. The close match between the actual change and the predicted one illustrates the validity of the proposed approach.

\subsection{Influence of domain size and spacing}
\begin{figure}[!h]
\centering
\begin{tabular}{ c | c | c | c | c | }
$d$ &
$r_0 = 0.9$ &
$r_0 = 1.0$ &
$r_0 = 1.1$ &
$r_0 = 1.2$ 
\\ \hline
$2.75$ &
\includegraphics[scale=0.08]{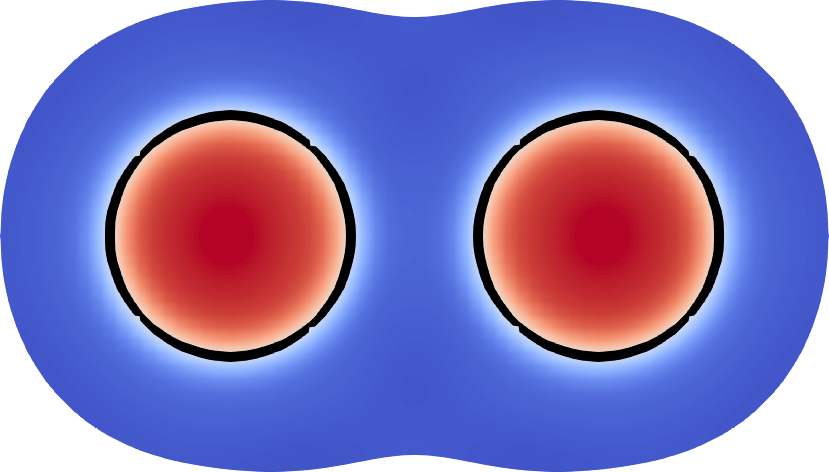} &
\includegraphics[scale=0.08]{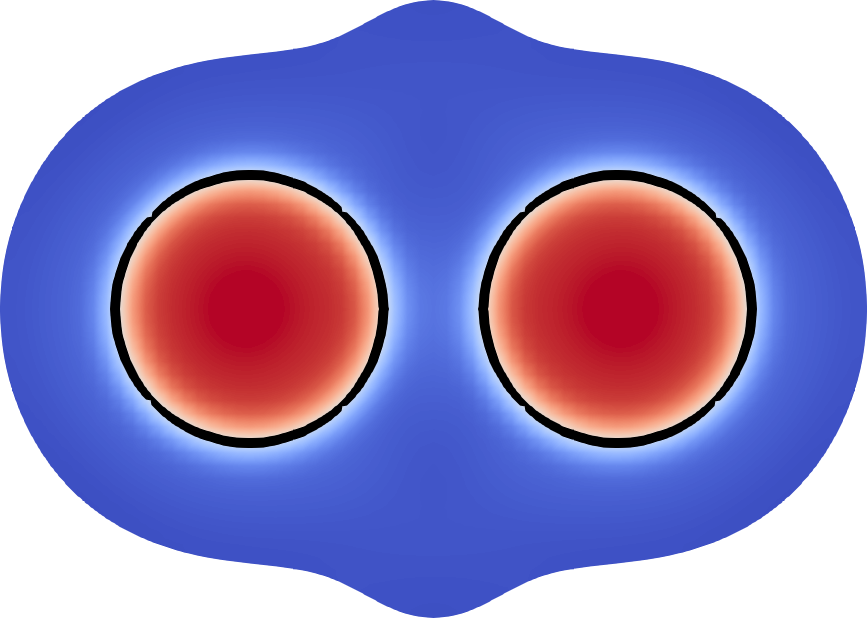} &
 &
\\ \hline
$3.00$ &
\includegraphics[scale=0.08]{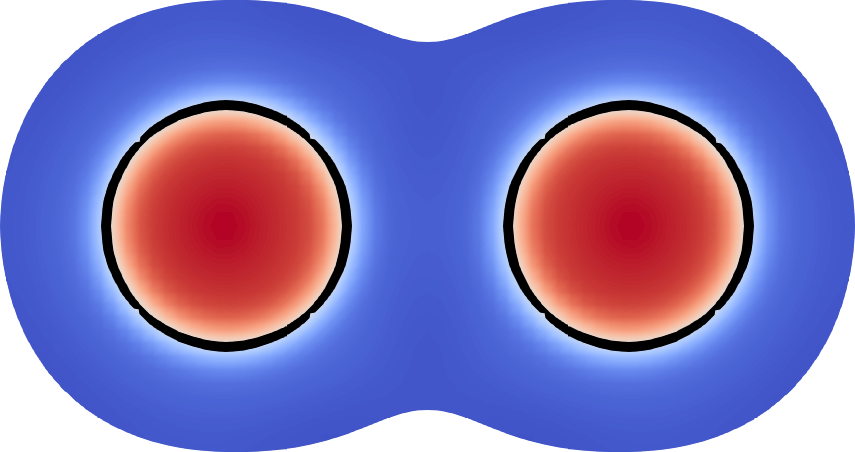} &
\includegraphics[scale=0.08]{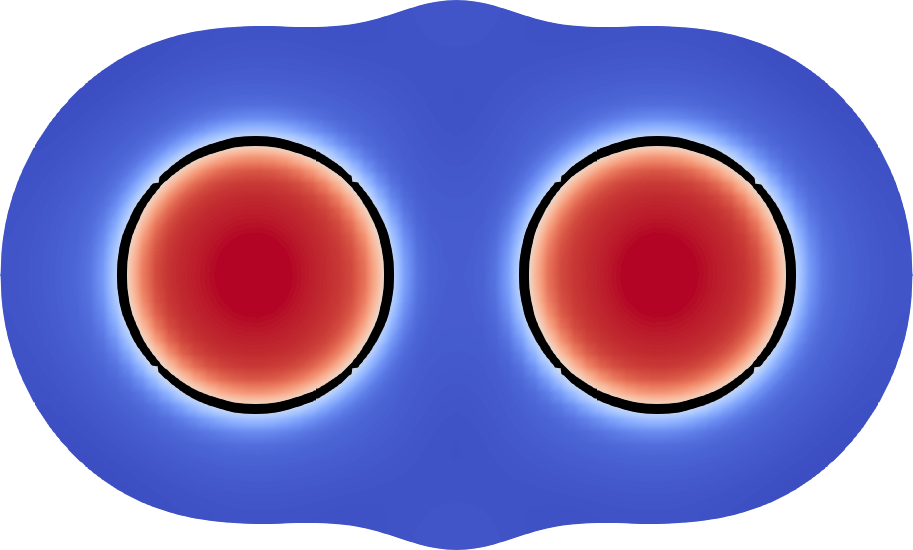} &
\includegraphics[scale=0.08]{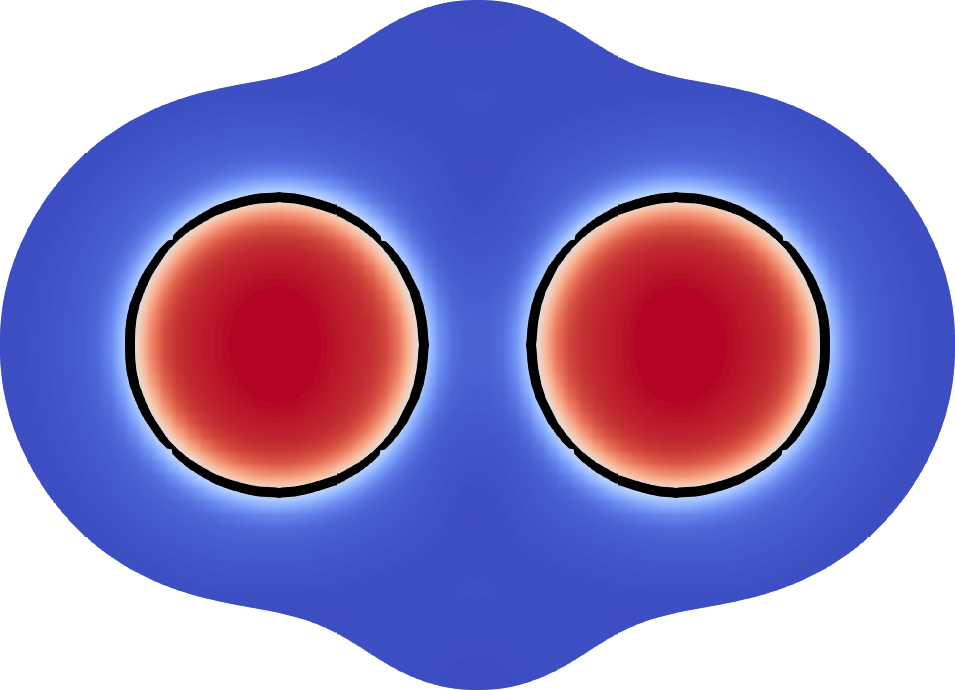} &
\\ \hline
$3.25$ &
\includegraphics[scale=0.08]{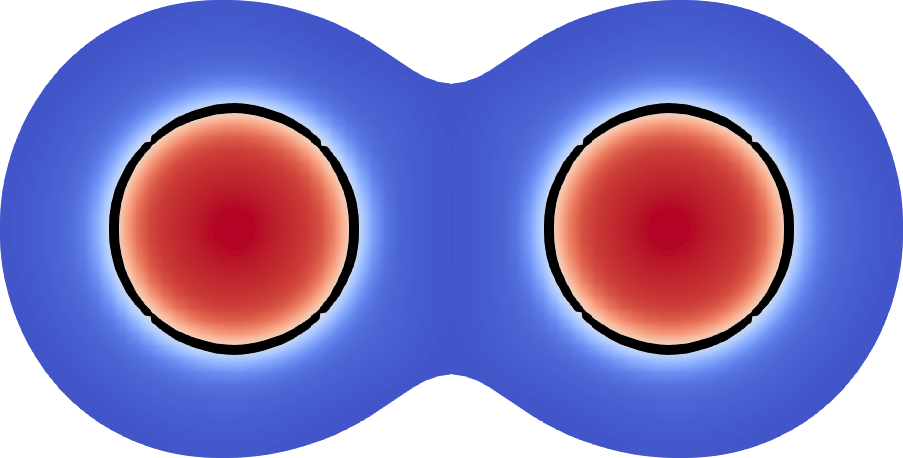} &
\includegraphics[scale=0.08]{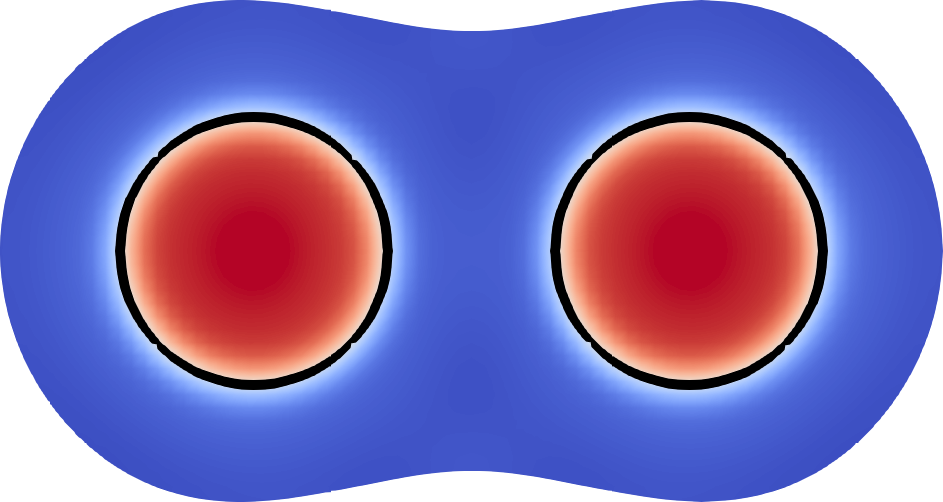} &
\includegraphics[scale=0.08]{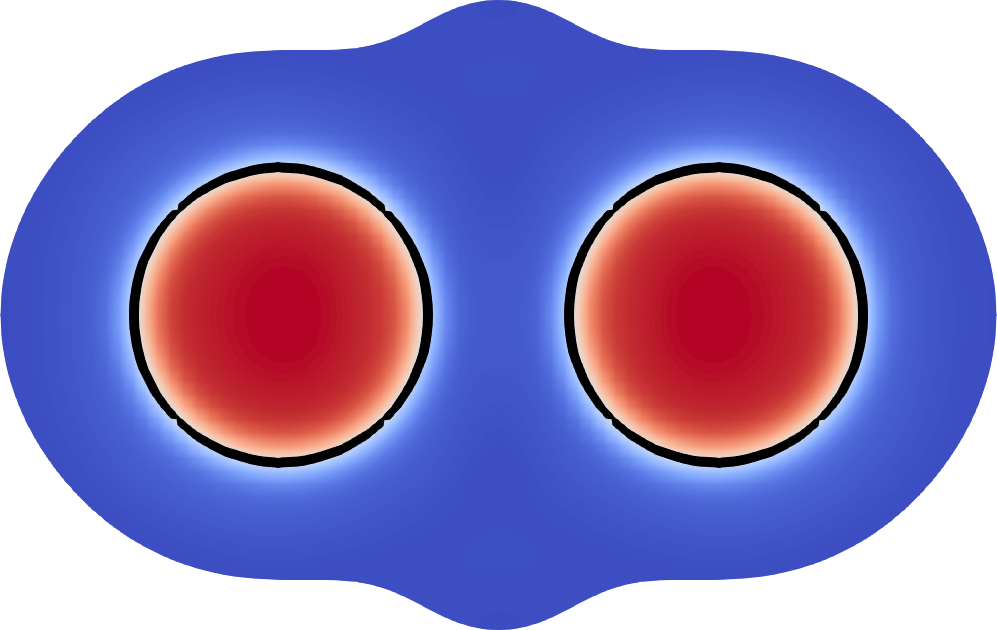} &
\includegraphics[scale=0.08]{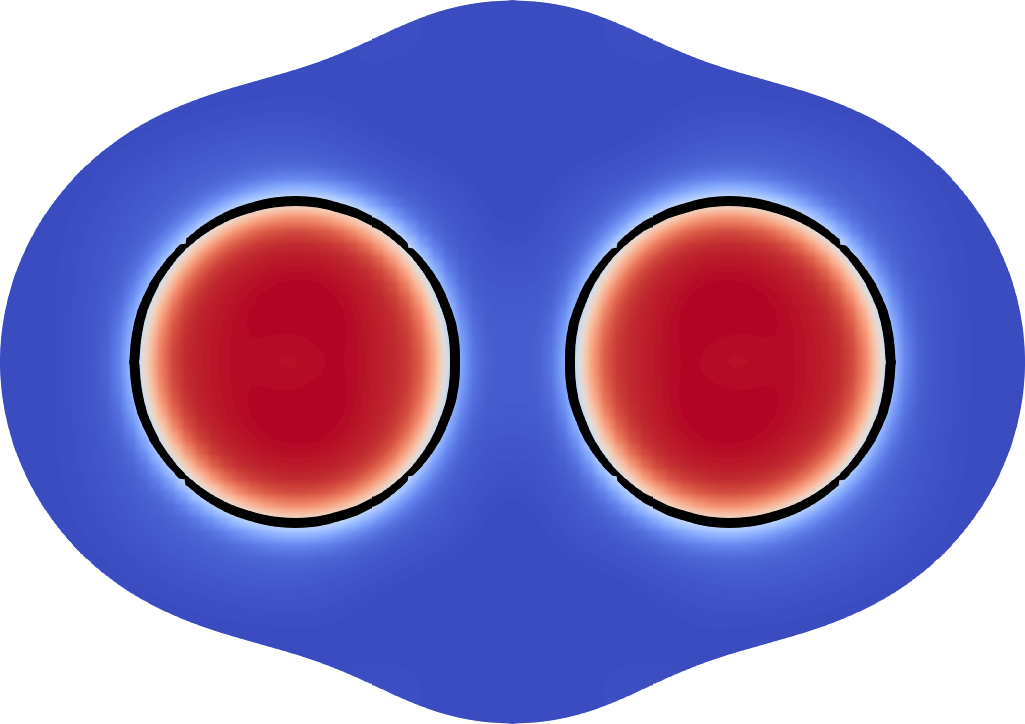}
\\ \hline
$3.50$ &
\includegraphics[scale=0.08]{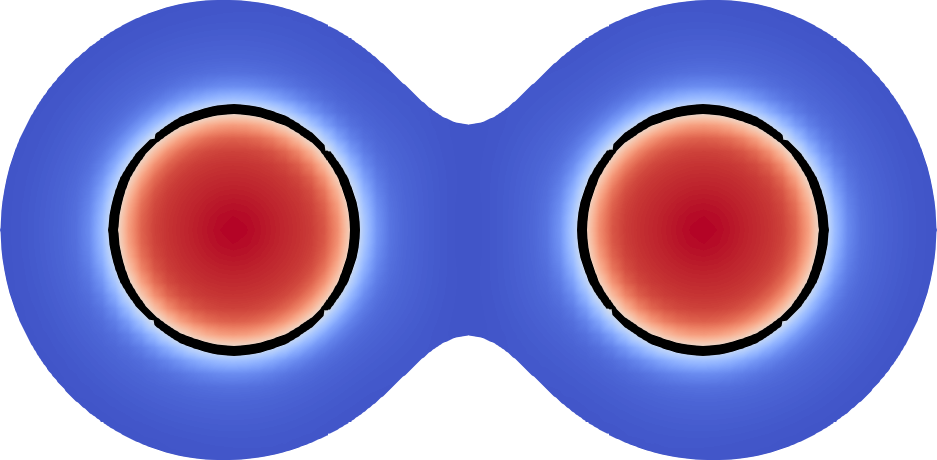} &
\includegraphics[scale=0.08]{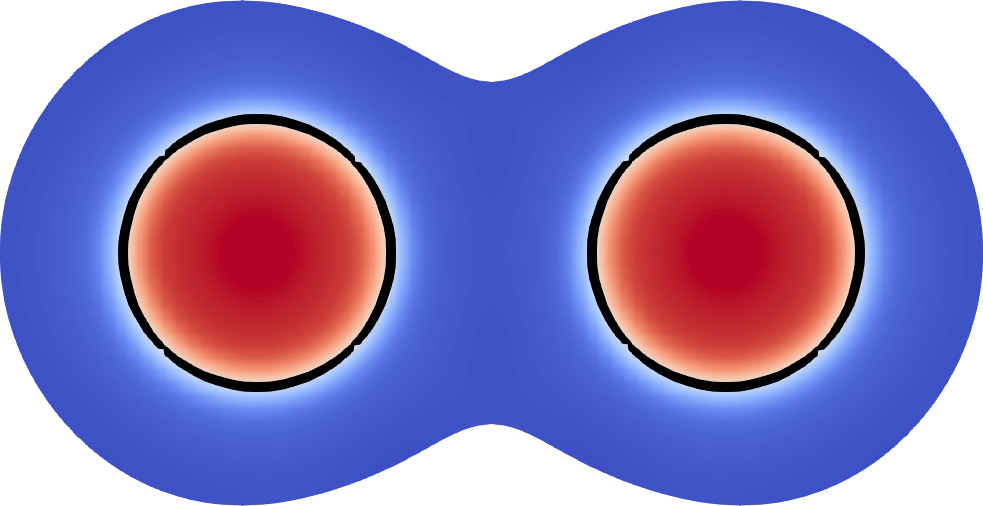} &
\includegraphics[scale=0.08]{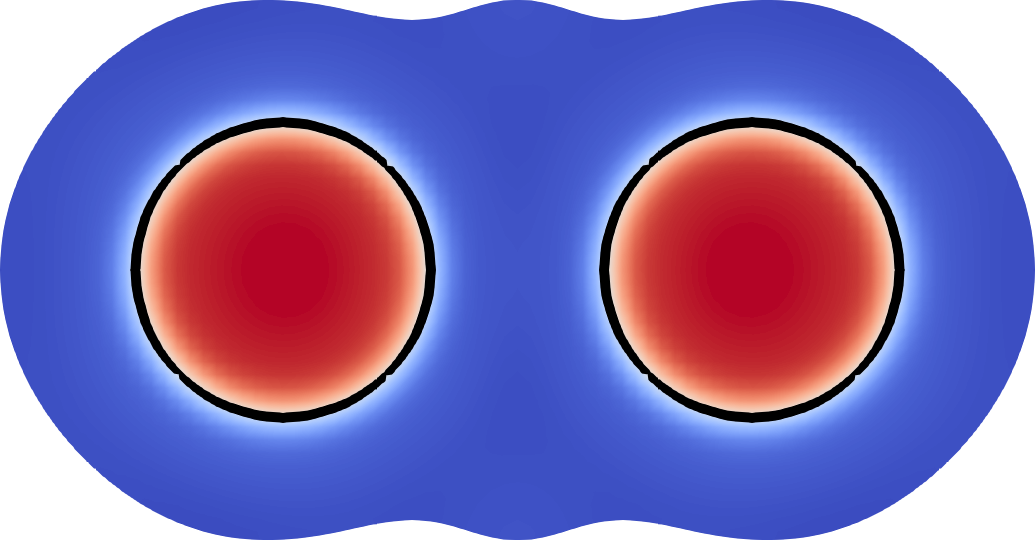} &
\includegraphics[scale=0.08]{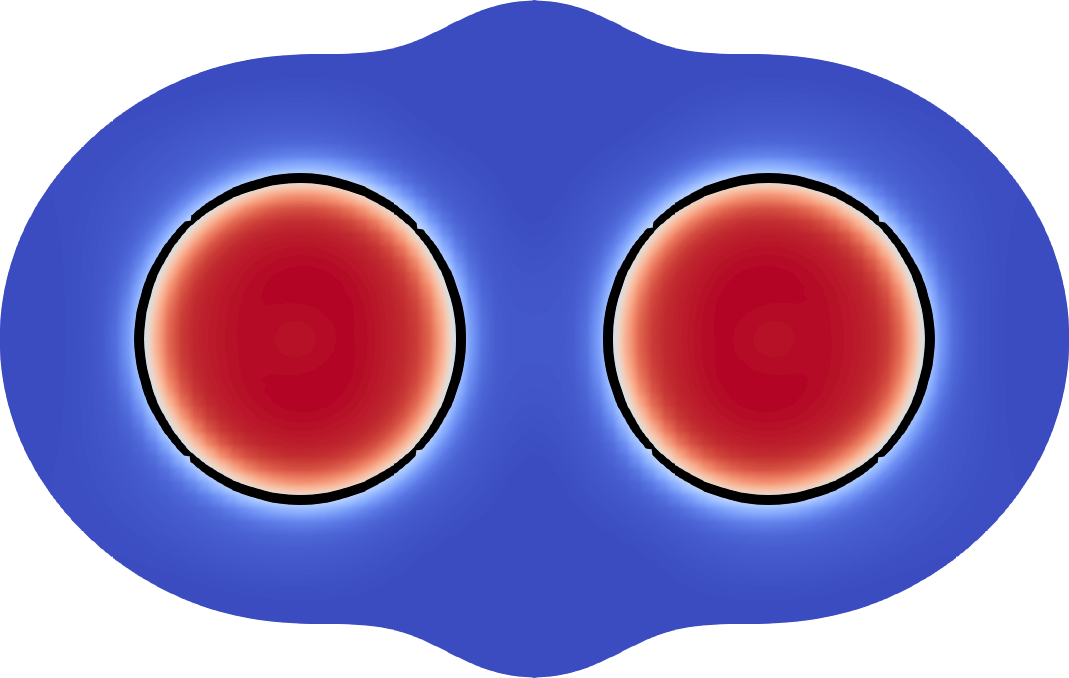}
\\ \hline
$3.75$ &
\includegraphics[scale=0.08]{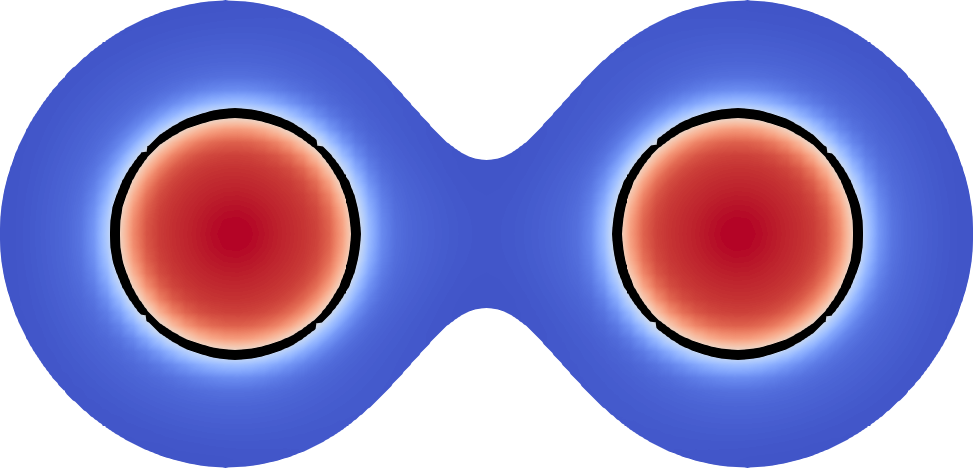} &
\includegraphics[scale=0.08]{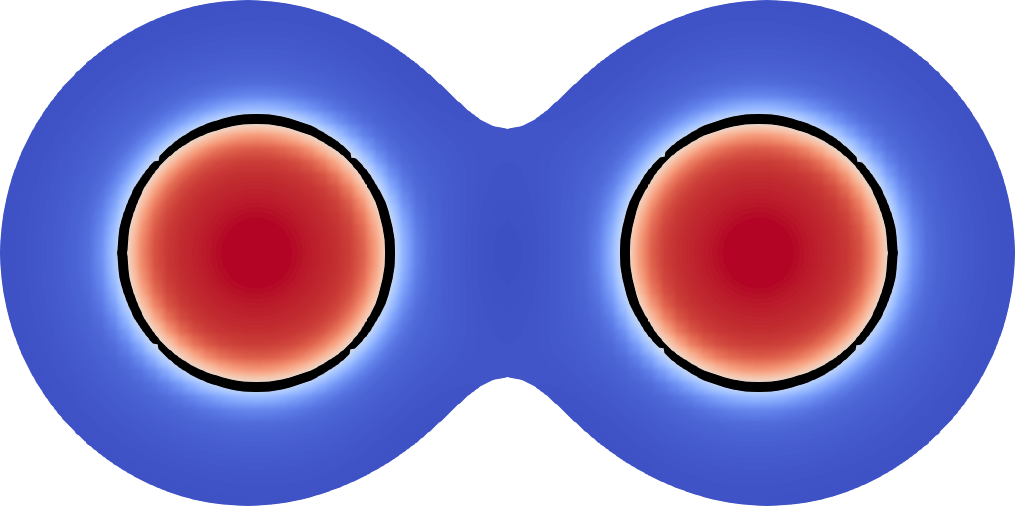} &
\includegraphics[scale=0.08]{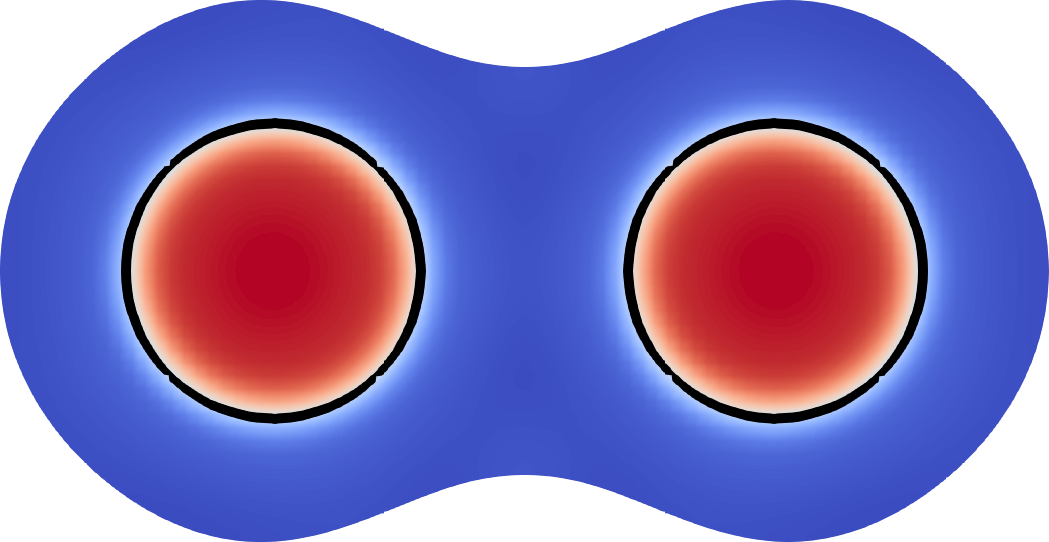} &
\includegraphics[scale=0.08]{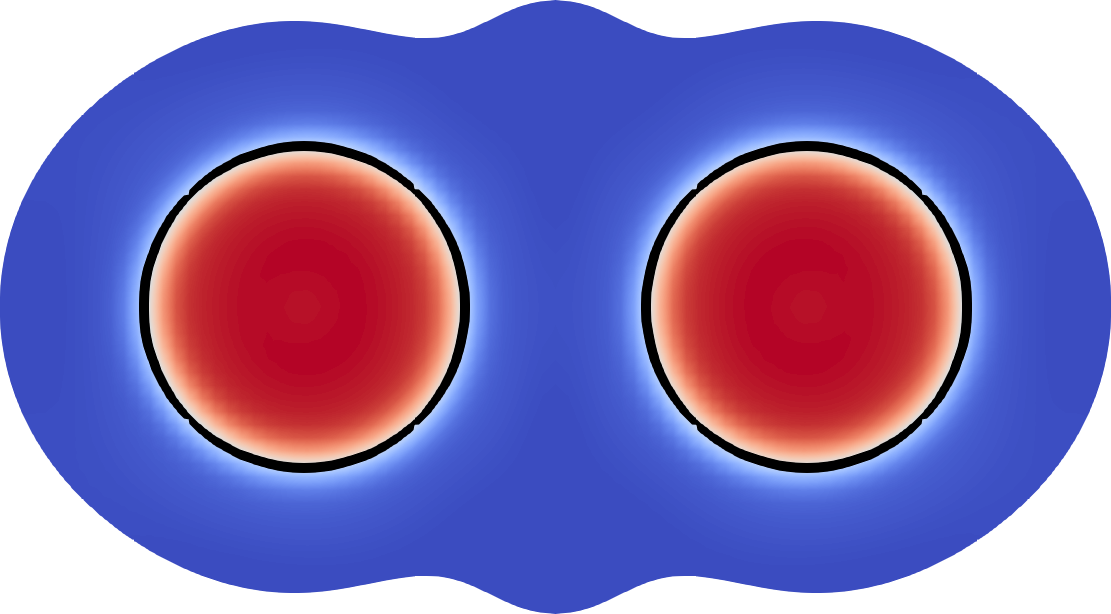}
\\ \hline
$4.00$ &
\includegraphics[scale=0.08]{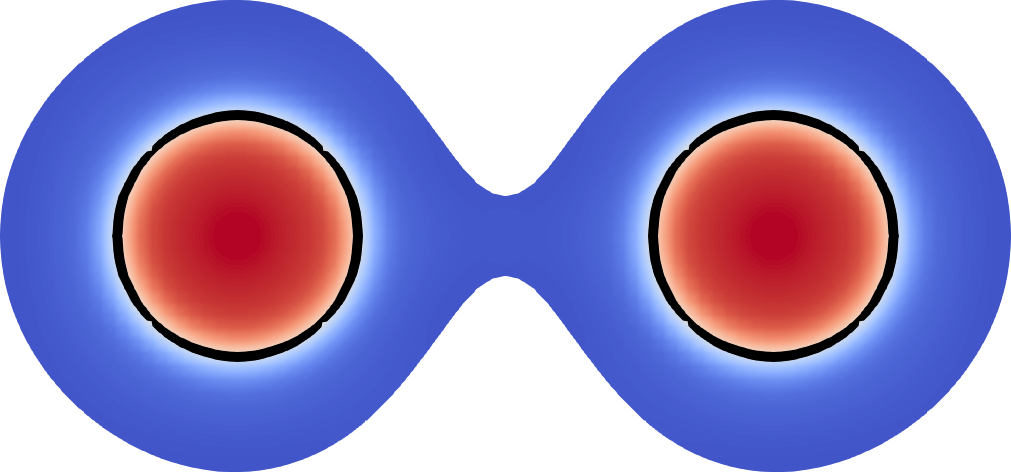} &
\includegraphics[scale=0.08]{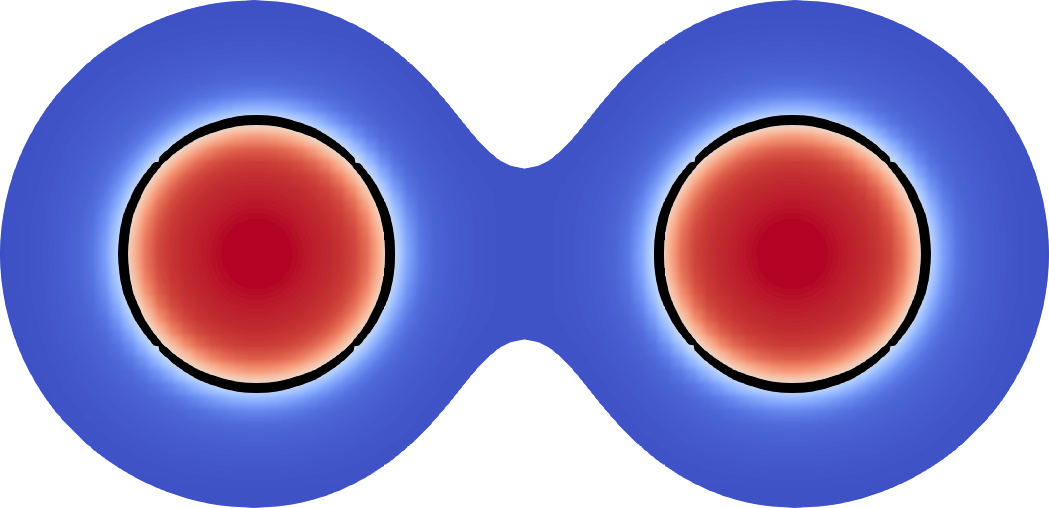} &
\includegraphics[scale=0.08]{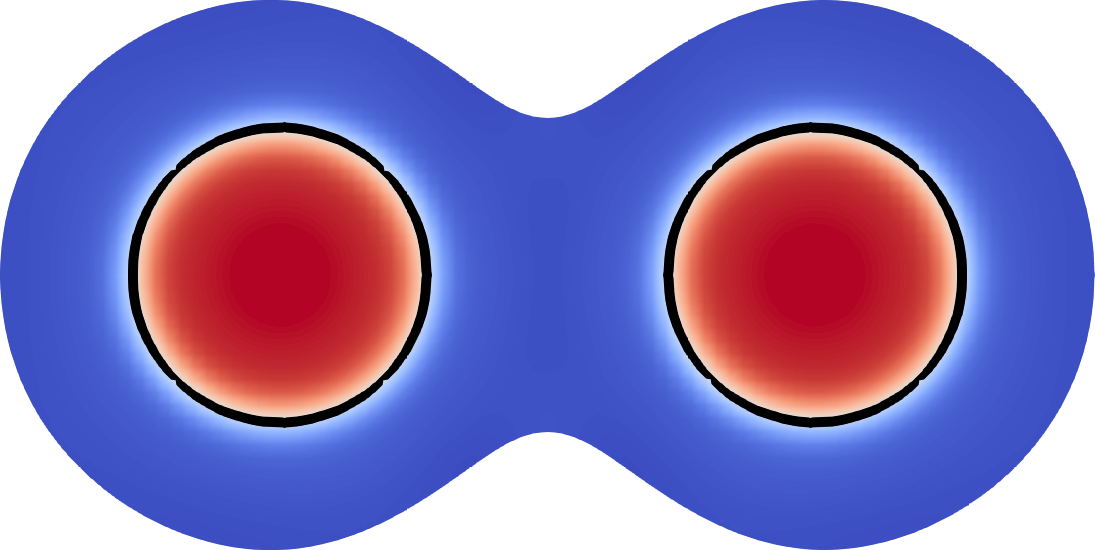} &
\includegraphics[scale=0.08]{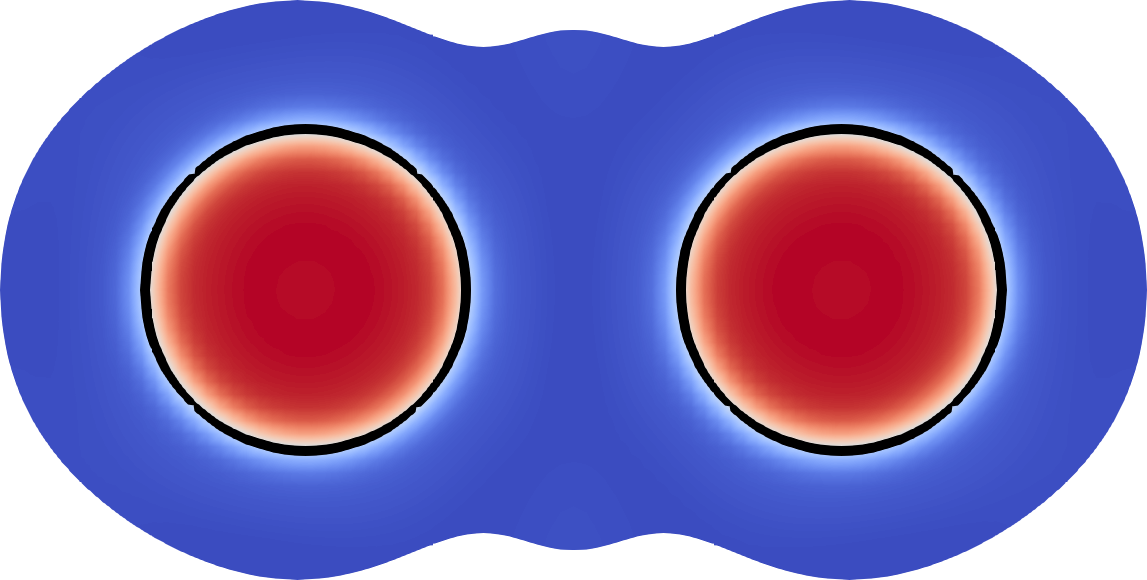}
\\ \hline
\end{tabular}
\caption{Confining masks for placement of two cylinders for different design parameters. Coloring shows the density configuration of the self-assembled polymer while the solid black line represent the target template.}
\label{fig:results:dsa:two}
\end{figure}

We start with a simple case of designing confining masks for the placement of two cylindrical domains formed by the minority component. Specifically, we investigate the influence of the domains' size and spacing on the resulting mask shapes. Figure \ref{fig:results:dsa:two} illustrates mask shapes obtained using the proposed optimization algorithm for cylindrical radii ranging from $r_0 = 0.9 R_g$ to $r_0 = 1.2 R_g$ and domain spacings ranging from $\Delta r = 2.75 R_g$ to $\Delta r = 4 R_g$, where $R_g$ is the radius of gyration of the block copolymer chains. The value of $\alpha = 0.1$ is used for curvature penalization.
The results indicate that the proposed method is successful for almost every case considered and produces confining geometries that results not only in accurate placement of domain centers but their overall shapes as well. In some cases, the algorithm is able to find a mask geometry that will guide the self-assembly towards the target design. However, those cases consider small well distances and/or large cylinders radii, which forces the cylindrical polymer domains to be placed too close to each other and eventually merge. Thus, it should be interpreted as the non-existence of solution in these cases rather than the deficiency of the method. 

\begin{figure}[!h]
\centering
\begin{subfigure}{0.18\textwidth}
\centering
\includegraphics[width=.9\textwidth]{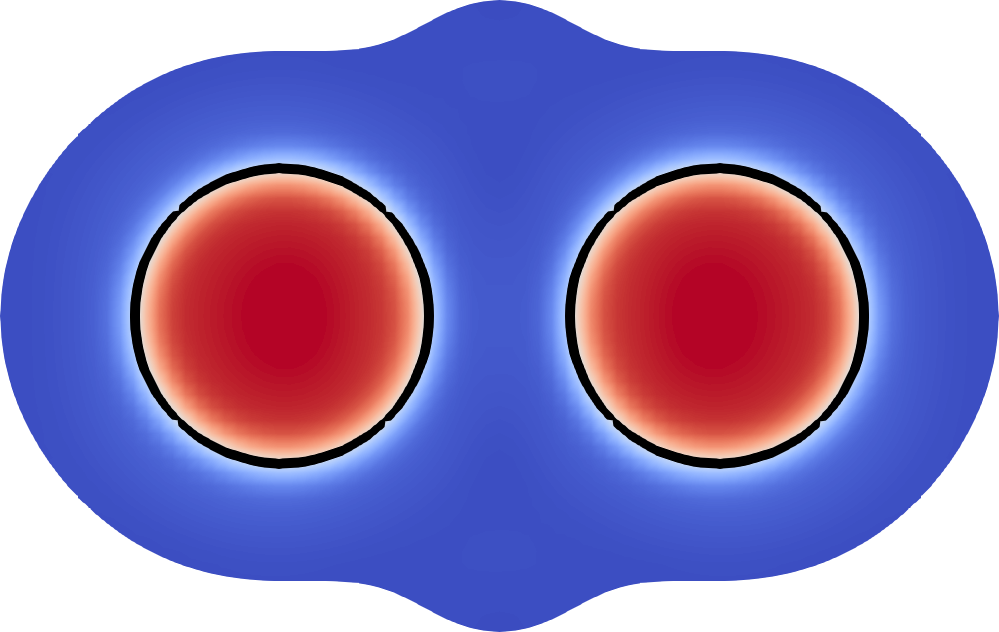}
\\
\medskip

\includegraphics[width=.99\textwidth]{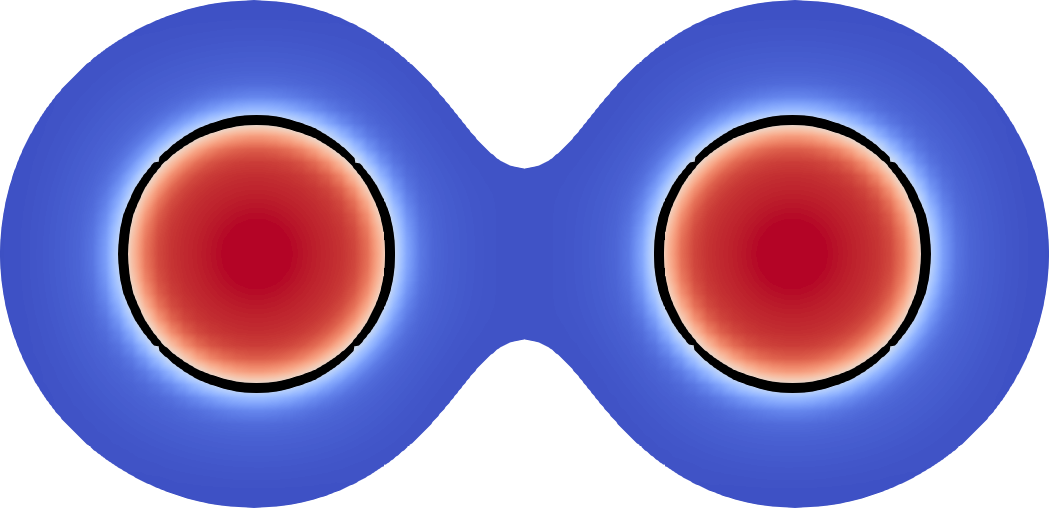}
\caption{$\alpha = 0.1$}
\end{subfigure}
\begin{subfigure}{0.18\textwidth}
\centering
\includegraphics[width=.9\textwidth]{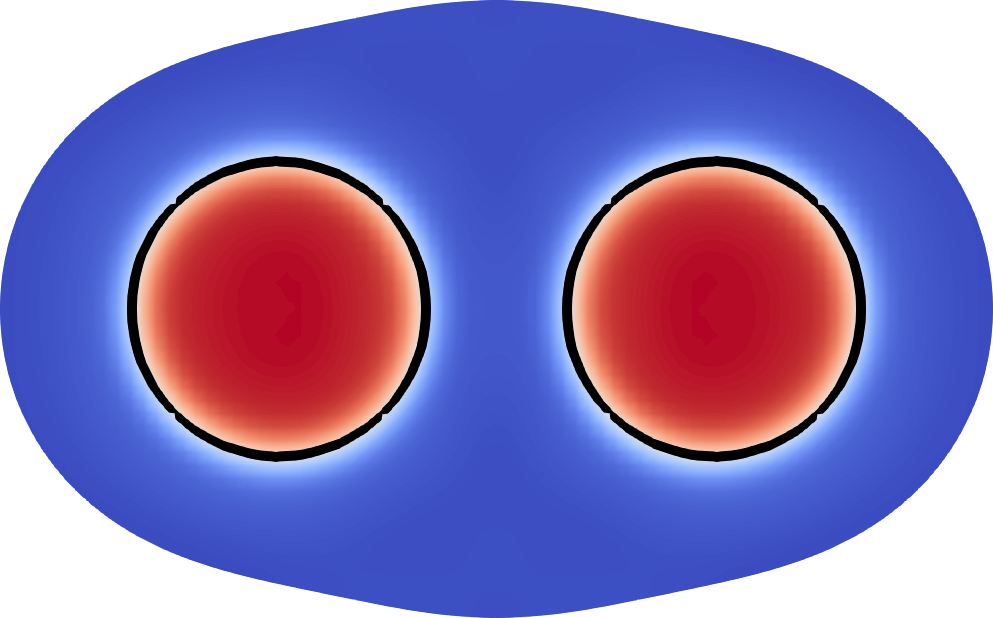}
\\
\medskip

\includegraphics[width=.99\textwidth]{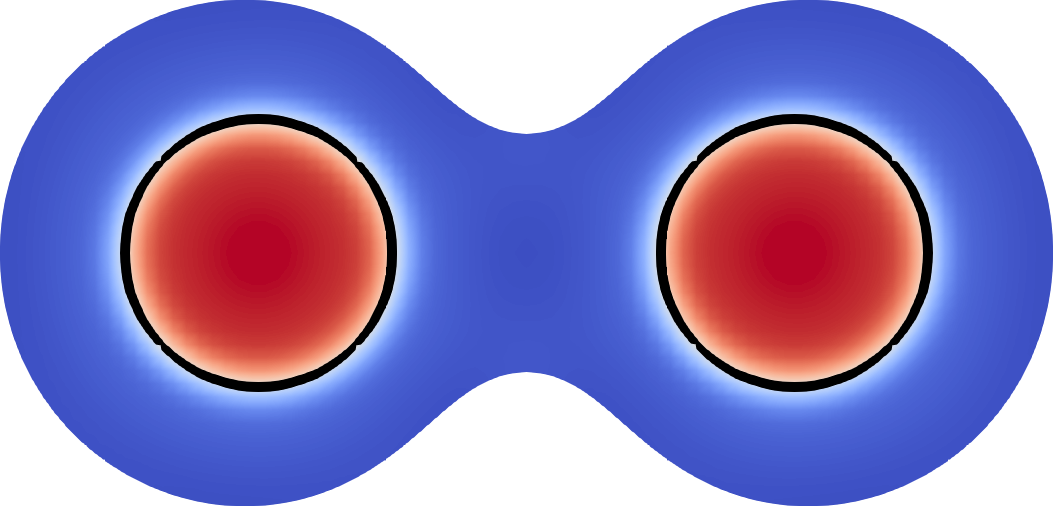}
\caption{$\alpha = 0.2$}
\end{subfigure}
\begin{subfigure}{0.18\textwidth}
\centering
\includegraphics[width=.9\textwidth]{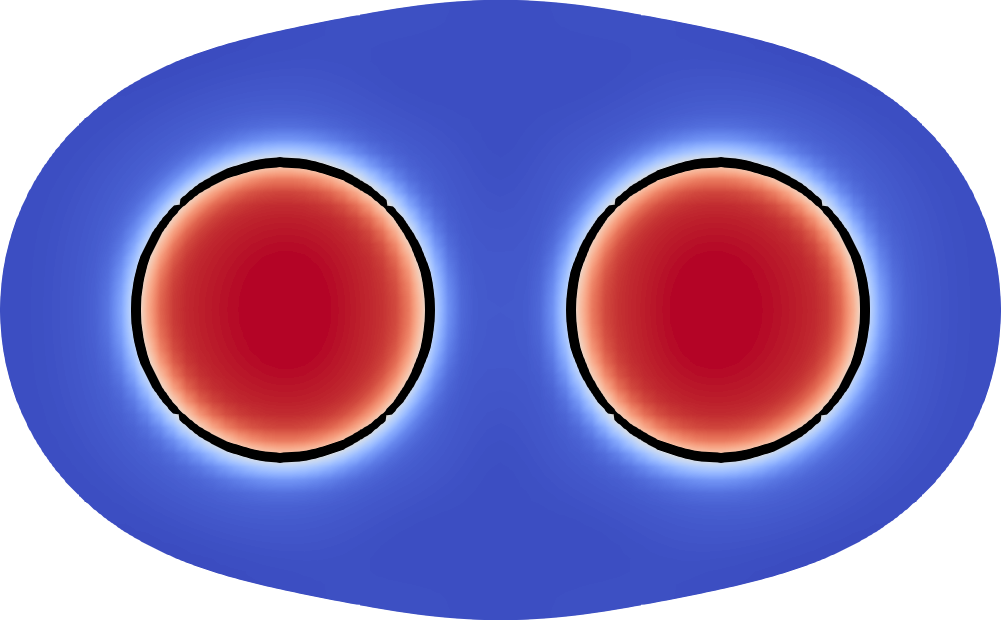}
\\
\medskip

\includegraphics[width=.99\textwidth]{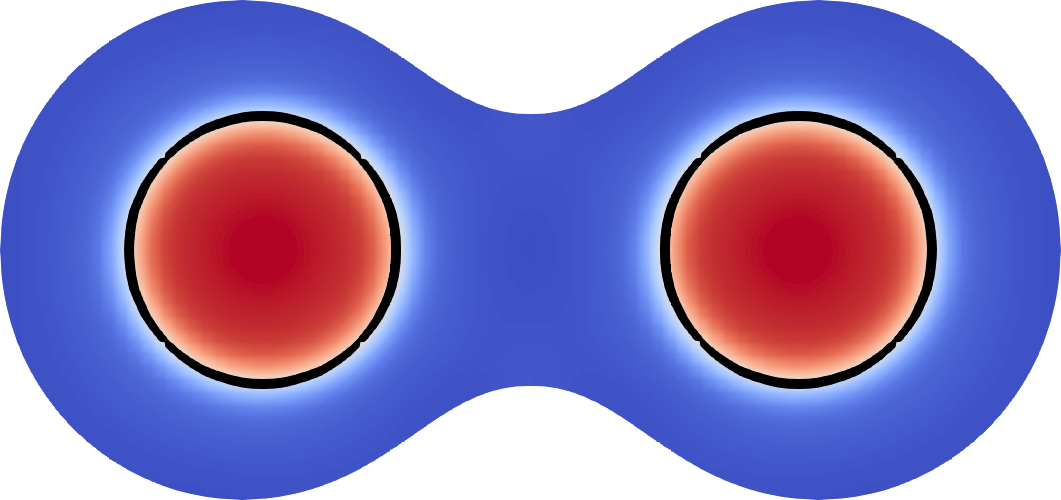}
\caption{$\alpha = 0.4$}
\end{subfigure}
\begin{subfigure}{0.18\textwidth}
\centering
\includegraphics[width=.9\textwidth]{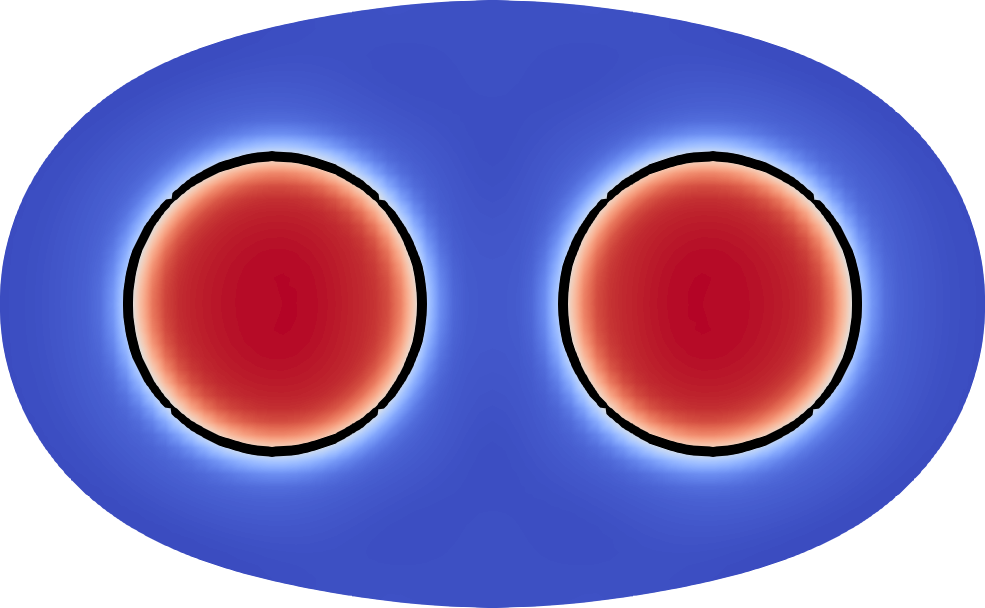}
\\
\medskip

\includegraphics[width=.99\textwidth]{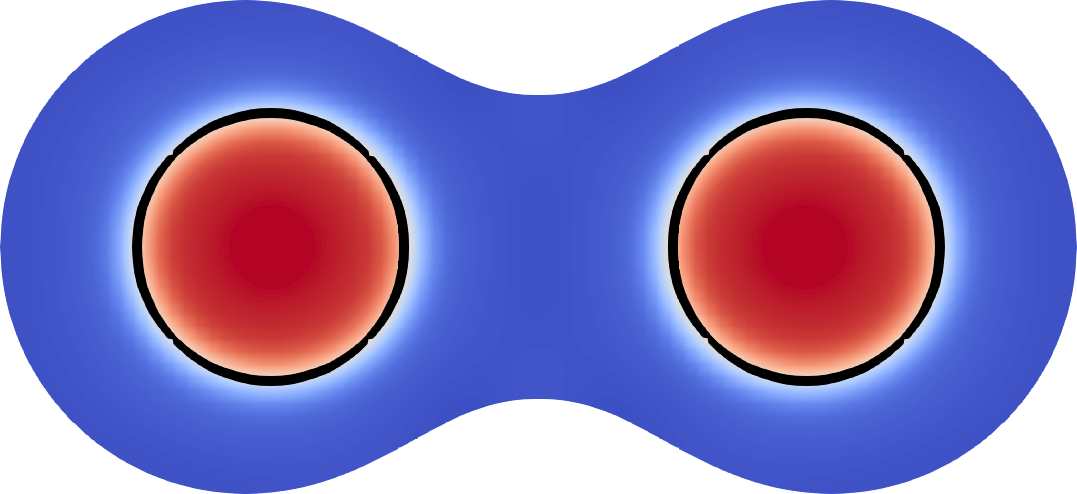}
\caption{$\alpha = 0.8$}
\end{subfigure}
\begin{subfigure}{0.18\textwidth}
\centering
\includegraphics[width=.9\textwidth]{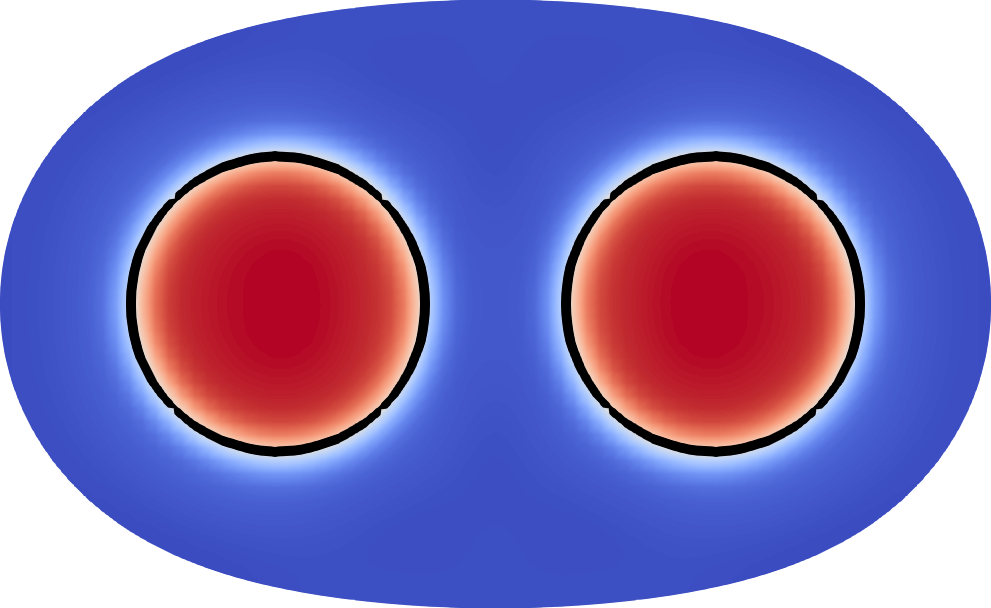}
\\
\medskip

\includegraphics[width=.99\textwidth]{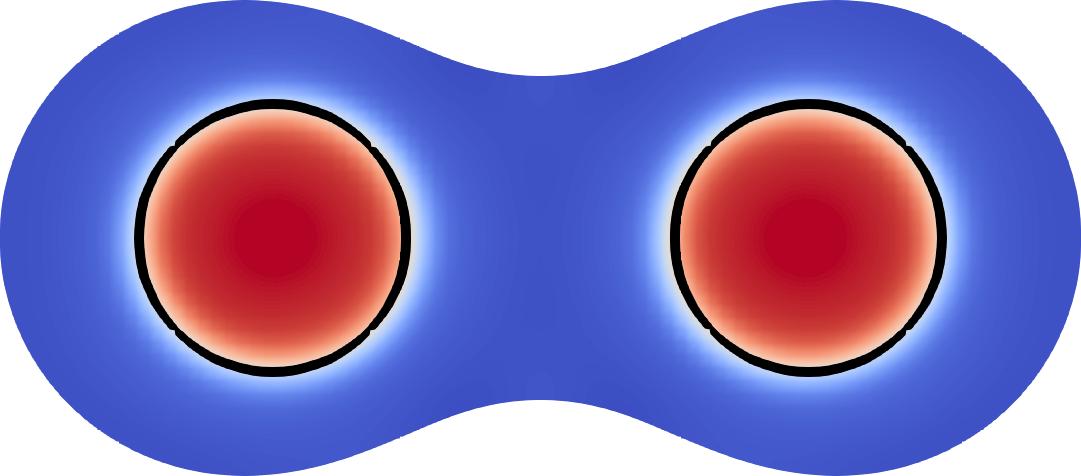}
\caption{$\alpha = 1.6$}
\end{subfigure}
\caption{Confining masks for two cylindrical domains placed $3 R_g$ (top row) and $4 R_g$ (bottom row) apart and varying the parameter $\alpha$ controlling the smoothness of the mask. Coloring shows the density configuration of the self-assembled polymer while the solid black line represent the target template.}
\label{fig:results:dsa:curvature}
\end{figure}

One can notice that depending on the target pattern geometry, the resulting confining masks can have either concave or convex features to accommodate for the polymer chains jammed between cylindrical domains. In either cases, such features have scales smaller than the target pattern, which makes such confining masks of a little practical value. However, we now show how the mask's smoothness can be controlled through adjusting the curvature penalization parameter $\alpha$. Specifically, we consider a case with concave features ($d = 4 R_g$, $r_0 = R_g$) and a case with convex features ($d = 3 R_g$, $r_0 = 1.1 R_g$). Figure \ref{fig:results:dsa:curvature} demonstrates the resulting confining mask in the cases where the curvature penalization parameter take the values $0.1$, $0.2$, $0.4$, $0.8$, and $1.6$.
As one can see, the proposed approach performs well under curvature constraints producing confining masks of desired smoothness that, at the same time, guide the polymer self-assembly very precisely to its target design.

\subsection{Influence of relative orientation}
Now we turn our attention to designing confining masks for the placement of a line of five cylinders making a turn at a specified angle. We consider angles ranging from $\theta = 60^{o}$ to $\theta = 150^{o}$. In addition, we investigate several cylinder spacing values ranging from $\Delta r = 3R_g$ to $\Delta r = 4R_g$. Figure \ref{fig:results:dsa:two} illustrates the resulting confining masks.
\begin{figure}[!h]
\centering
\begin{tabular}{ c | c | c | c | c | }
$d$ &
$60^o$ &
$90^o$ &
$120^o$ &
$150^o$ \\
\hline
$3.00$ &
\includegraphics[scale=0.07]{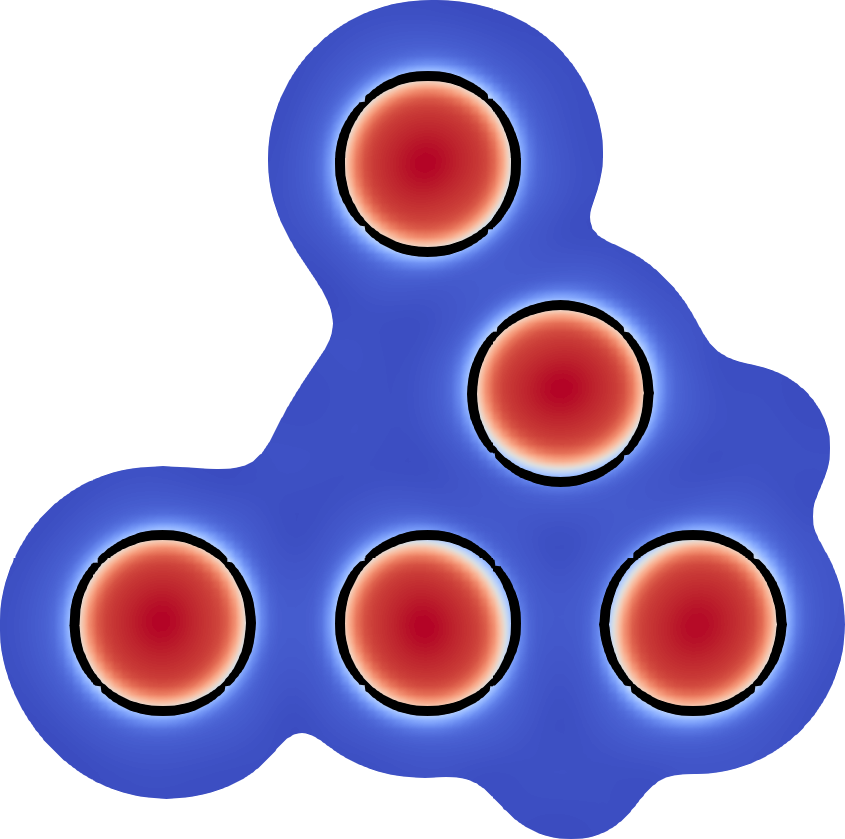} &
\includegraphics[scale=0.07]{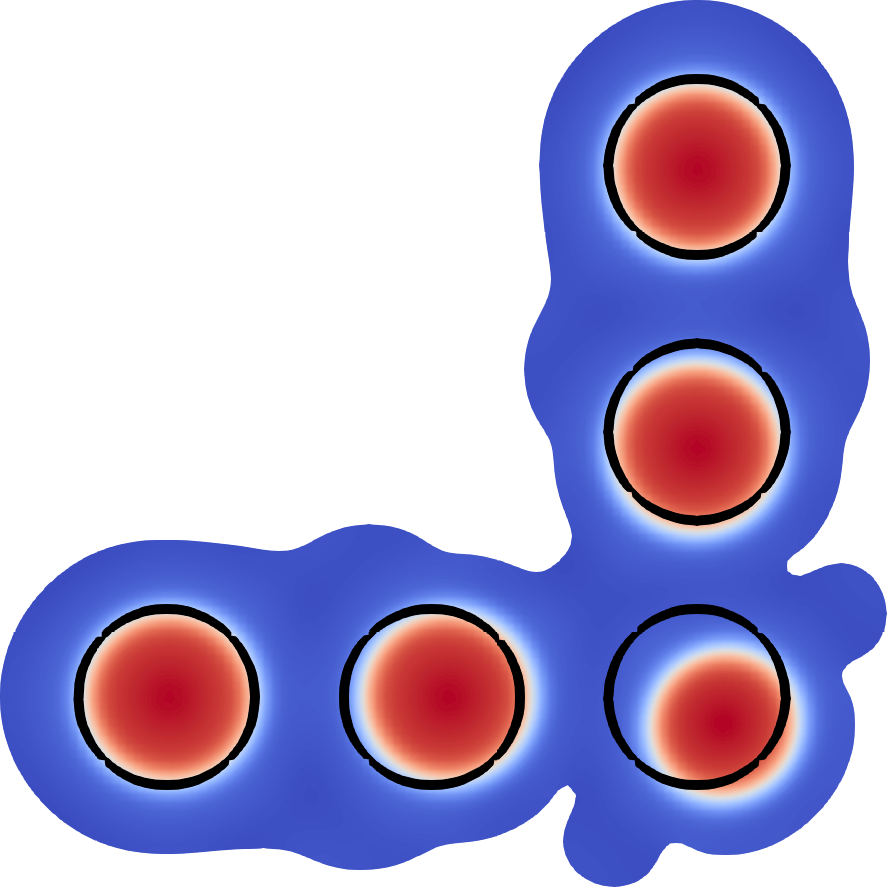} &
\includegraphics[scale=0.07]{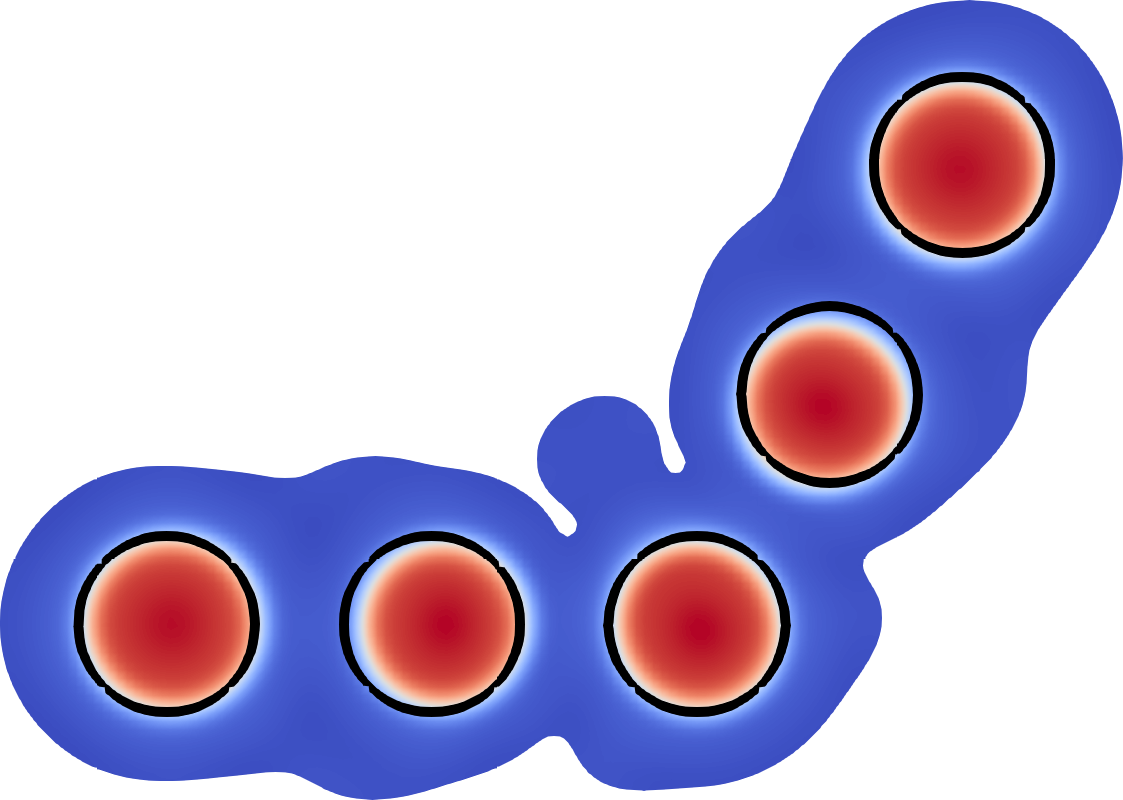} &
\includegraphics[scale=0.07]{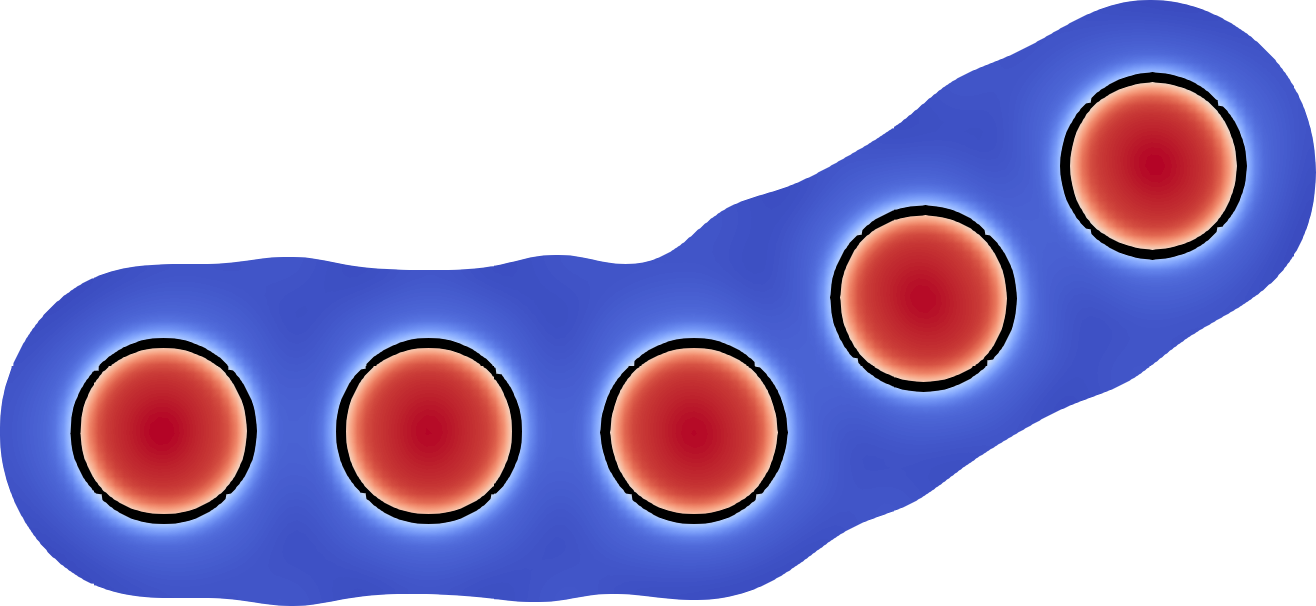} 
\\ \hline
$3.25$ &
\includegraphics[scale=0.07]{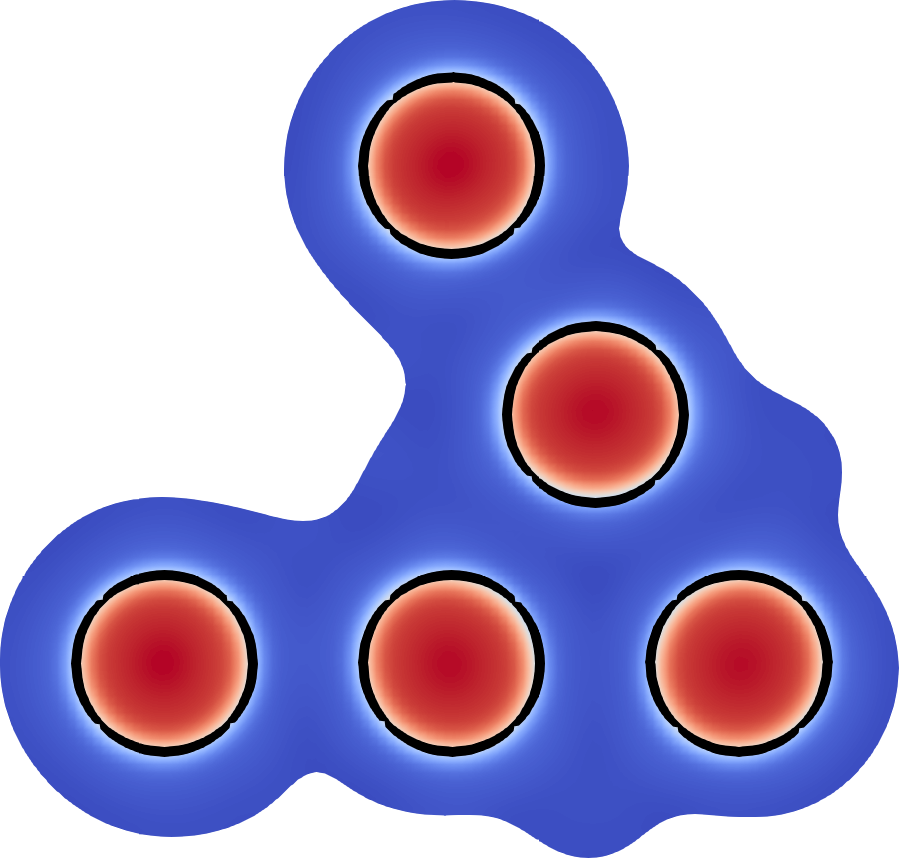} &
\includegraphics[scale=0.07]{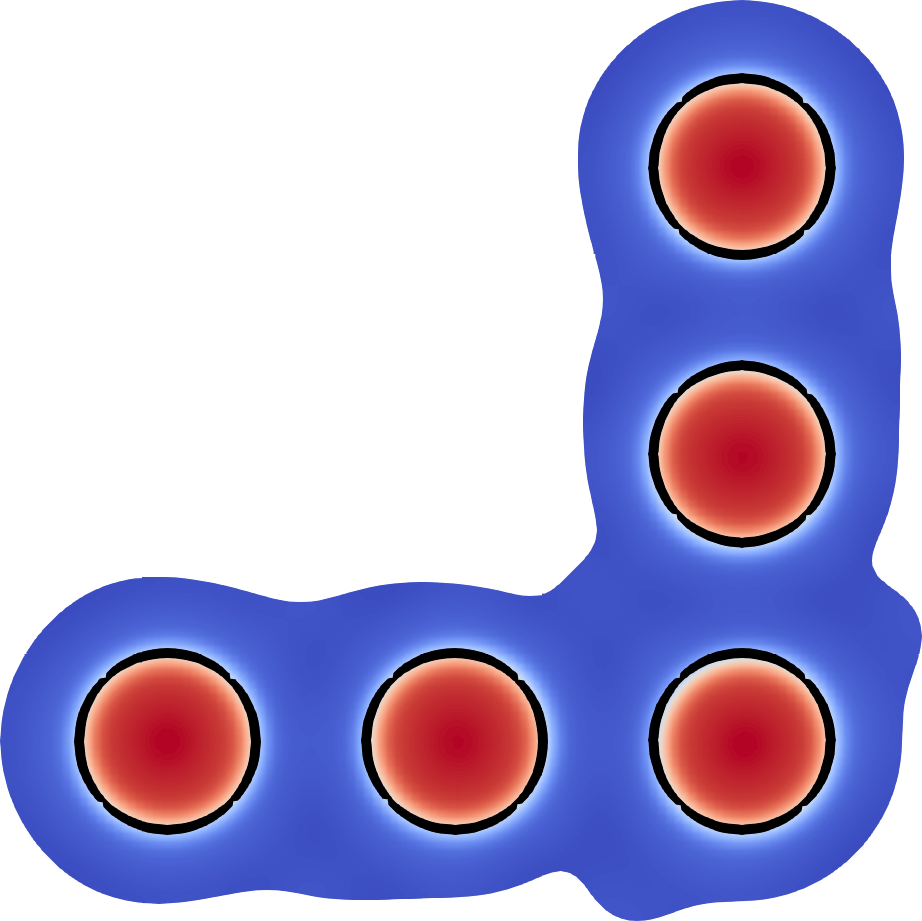} &
\includegraphics[scale=0.07]{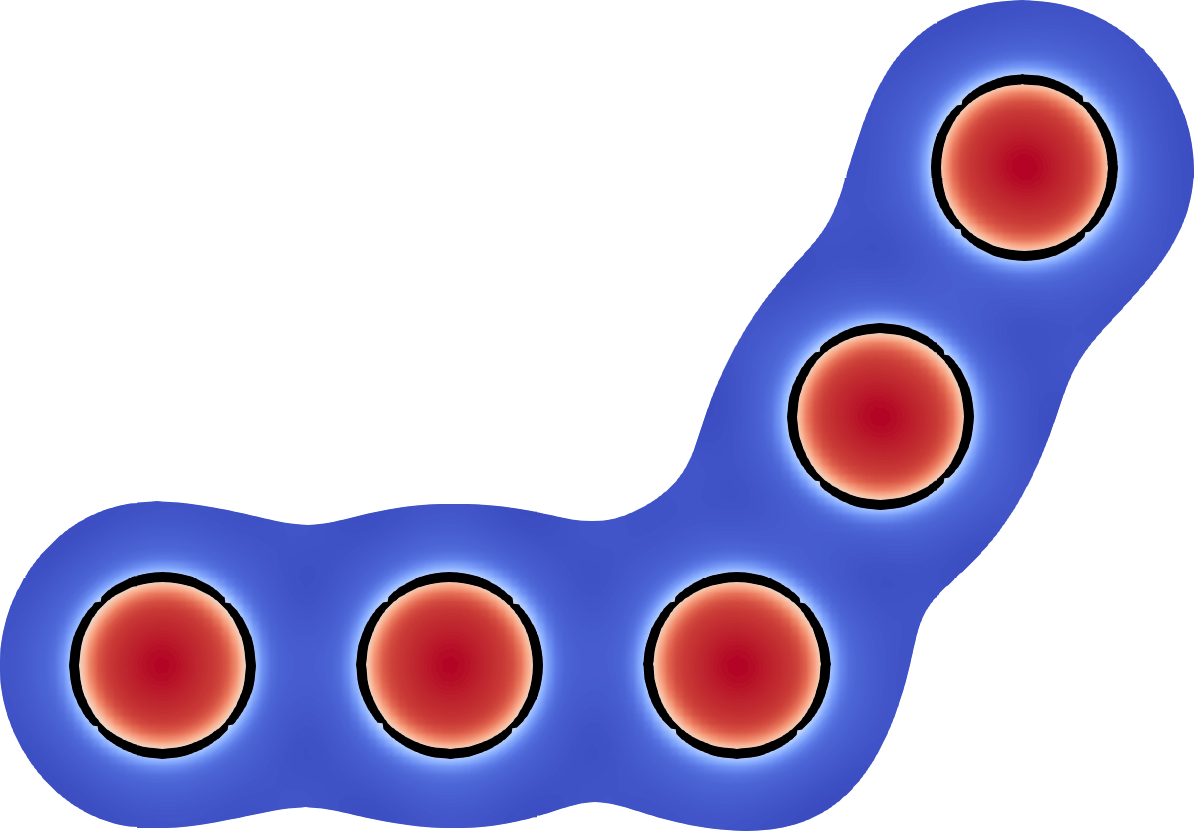} &
\includegraphics[scale=0.07]{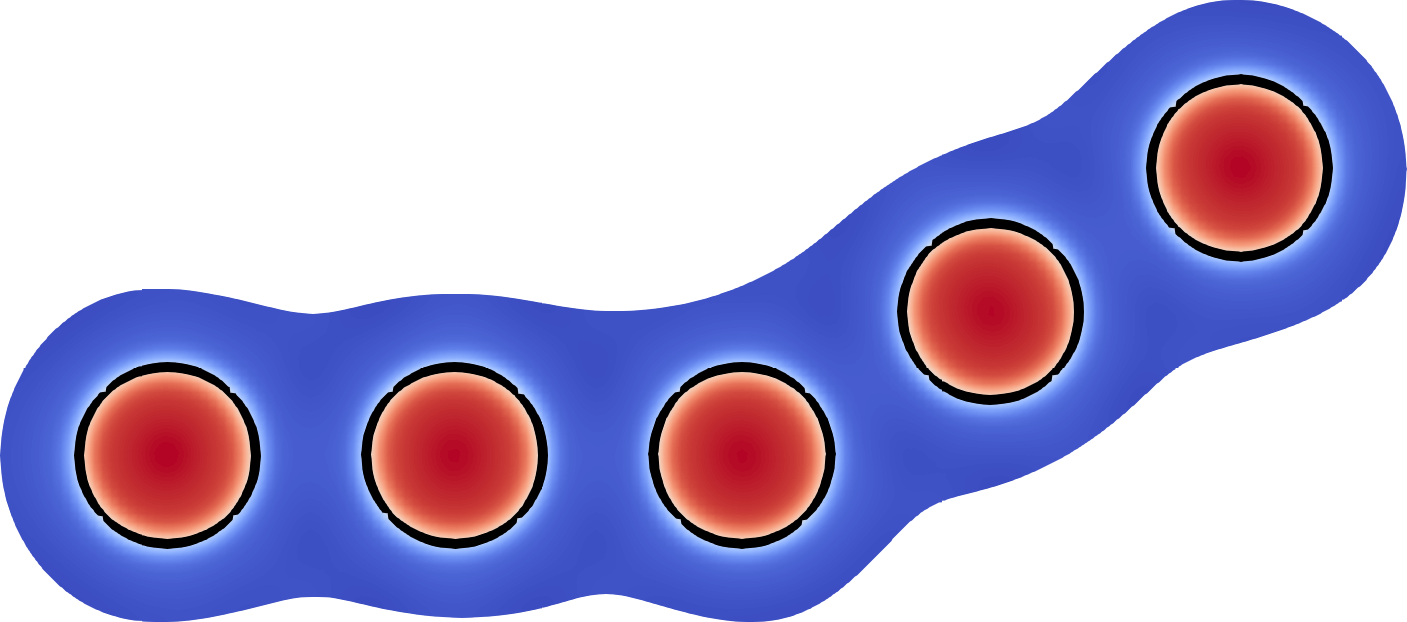} 
\\ \hline
$3.50$ &
\includegraphics[scale=0.07]{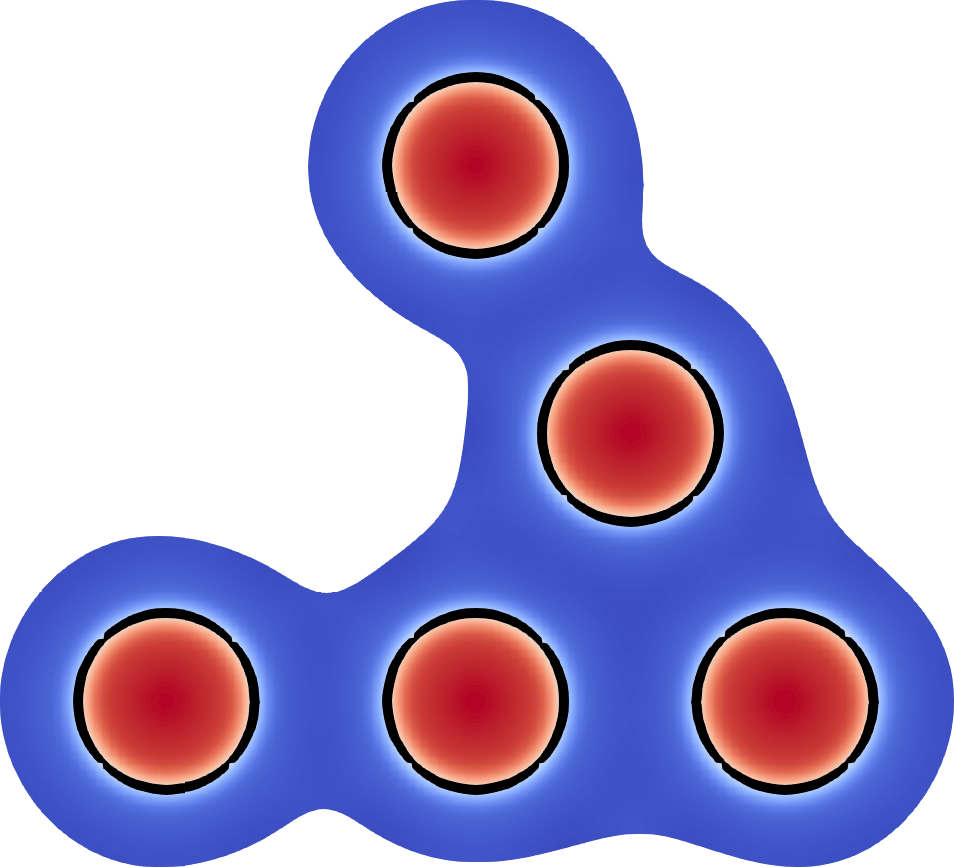} &
\includegraphics[scale=0.07]{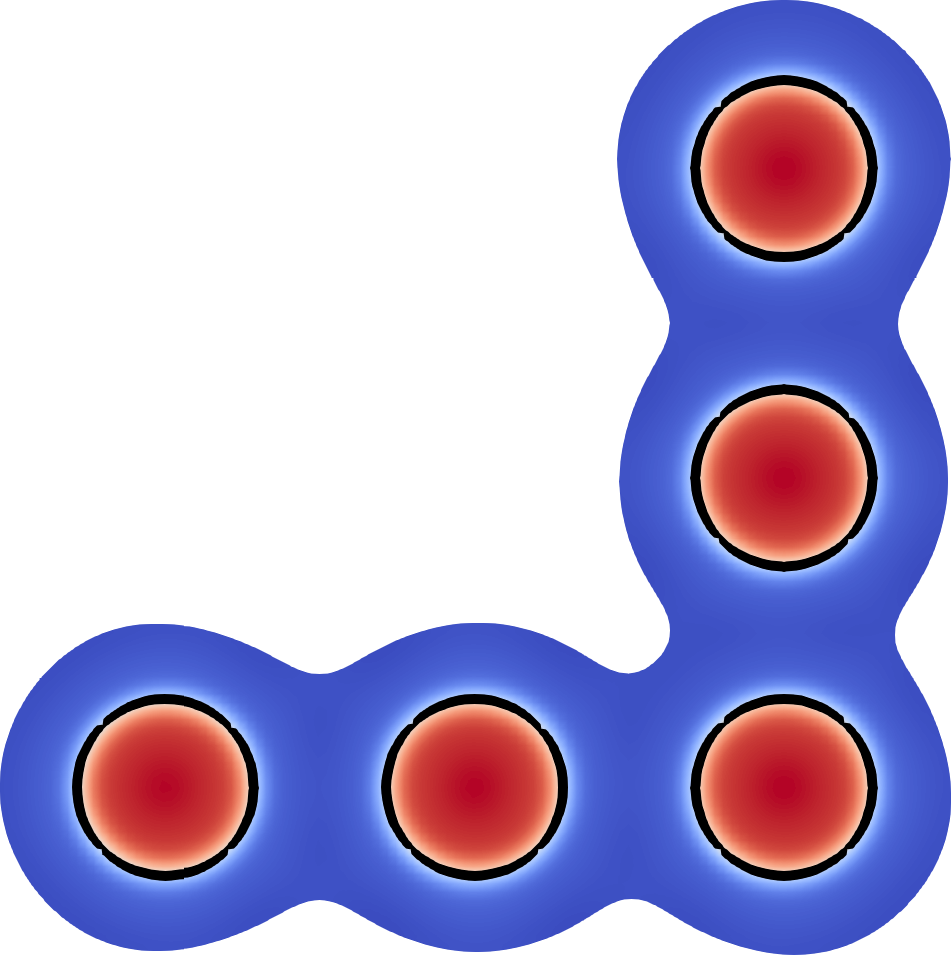} &
\includegraphics[scale=0.07]{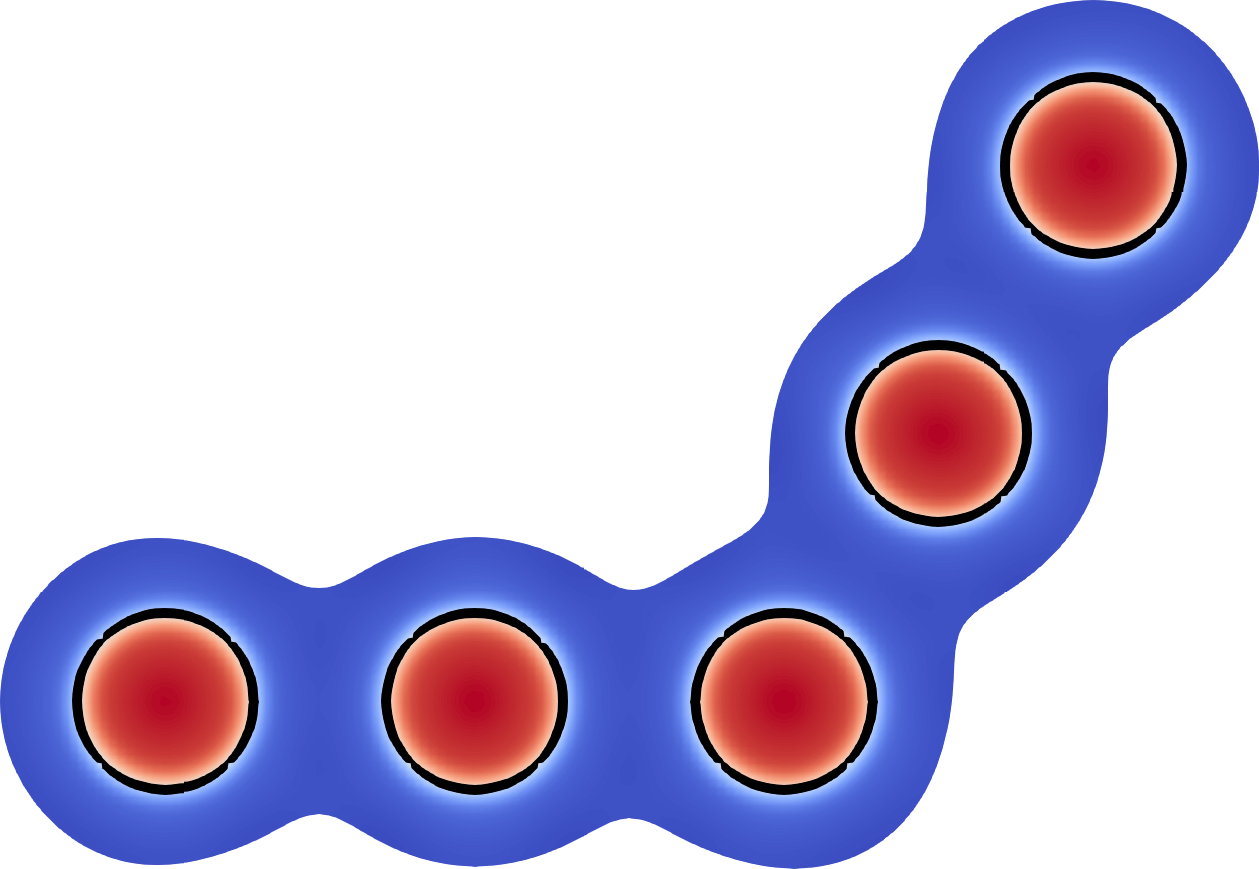} &
\includegraphics[scale=0.07]{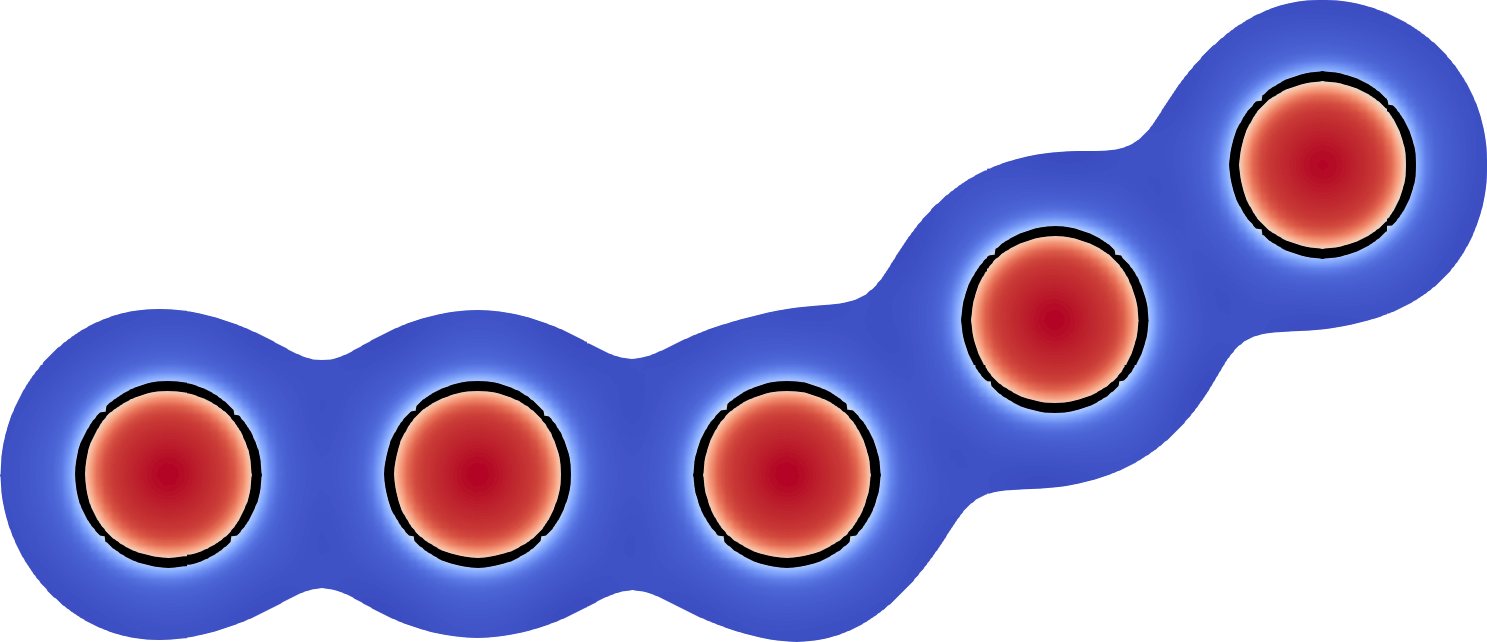} 
\\ \hline
$3.75$ &
\includegraphics[scale=0.07]{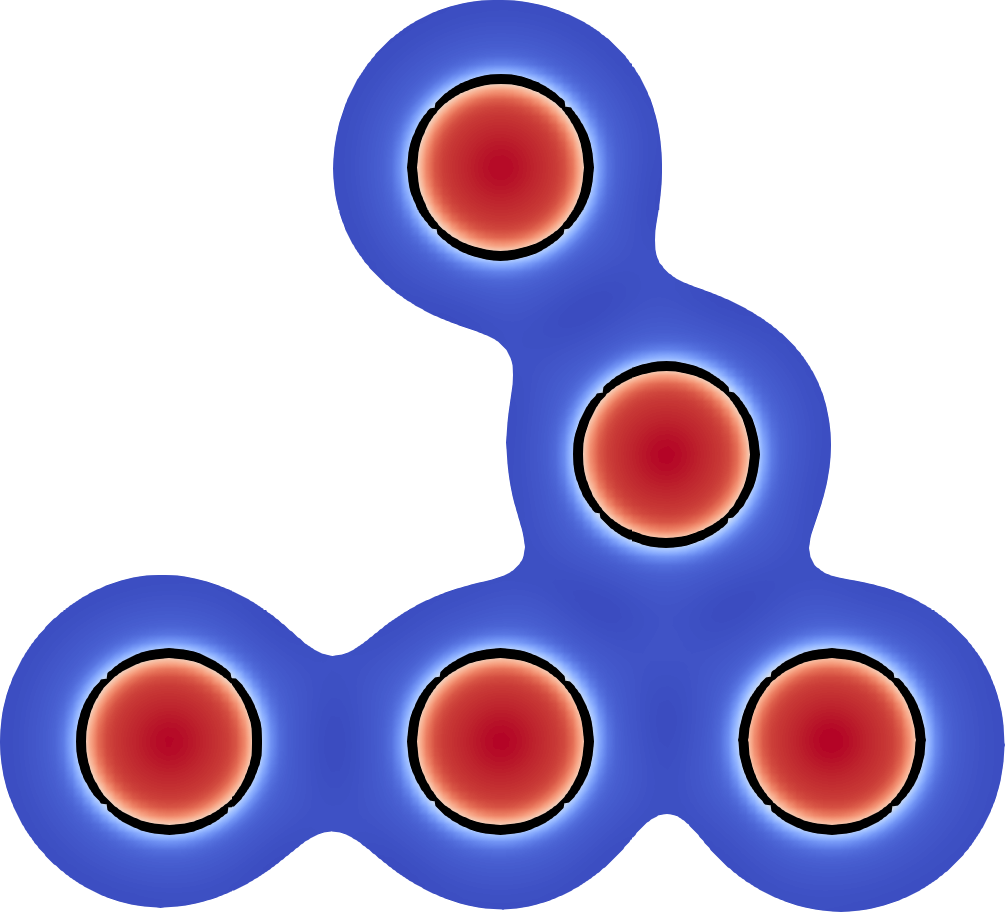} &
\includegraphics[scale=0.07]{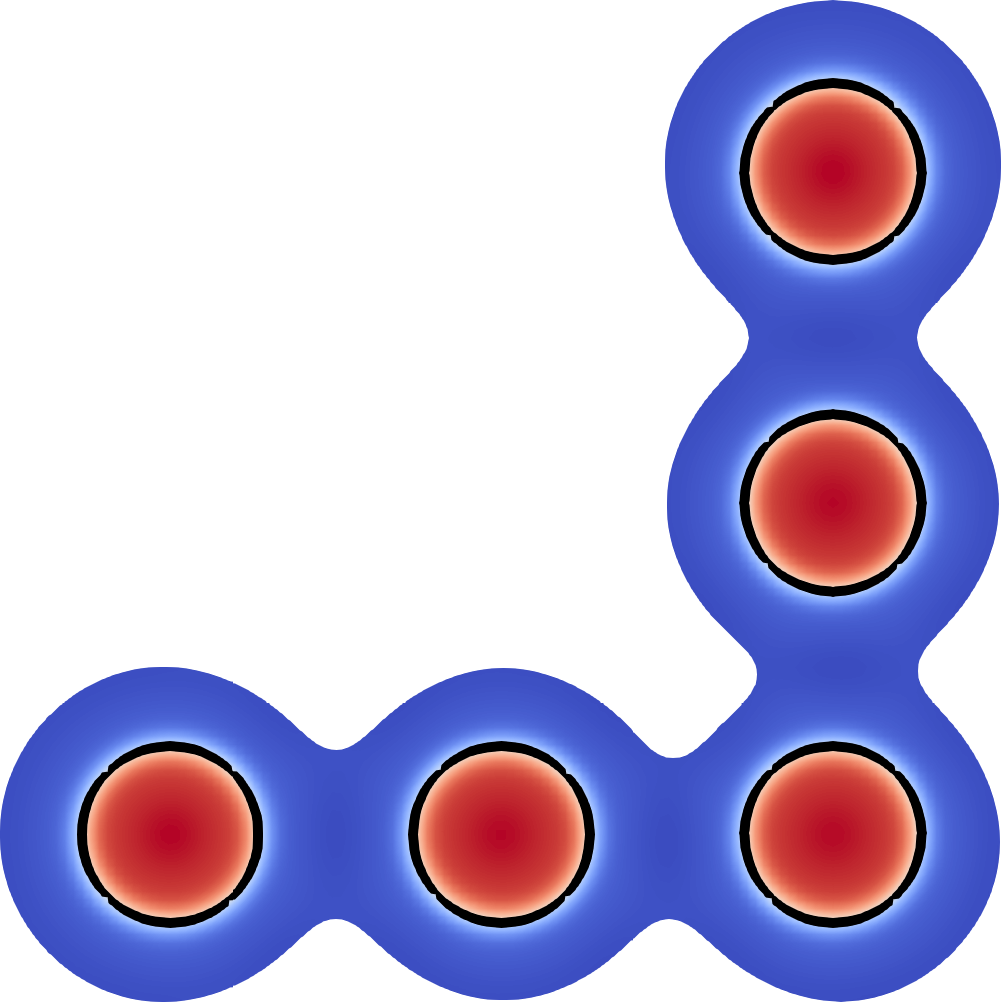} &
\includegraphics[scale=0.07]{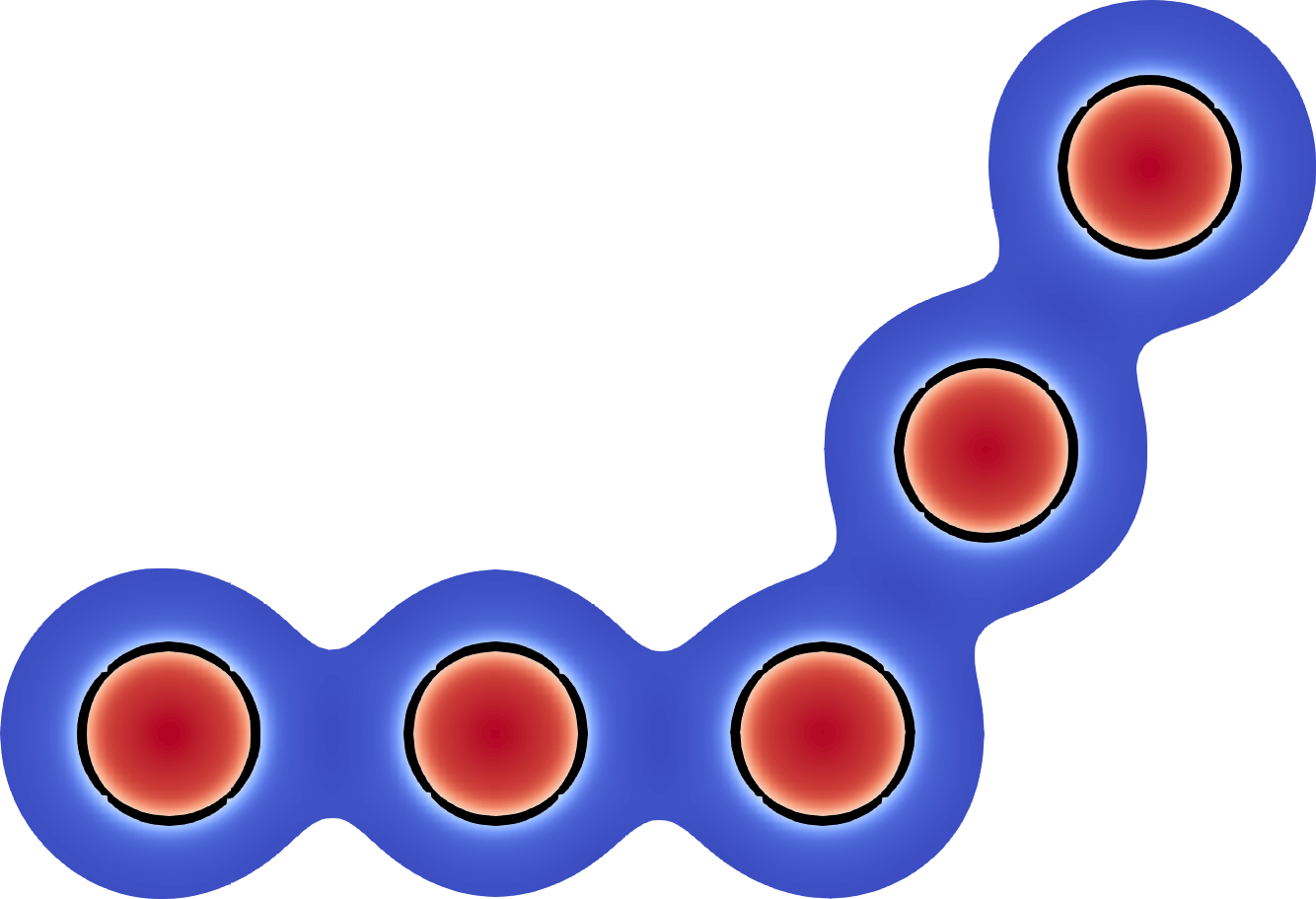} &
\includegraphics[scale=0.07]{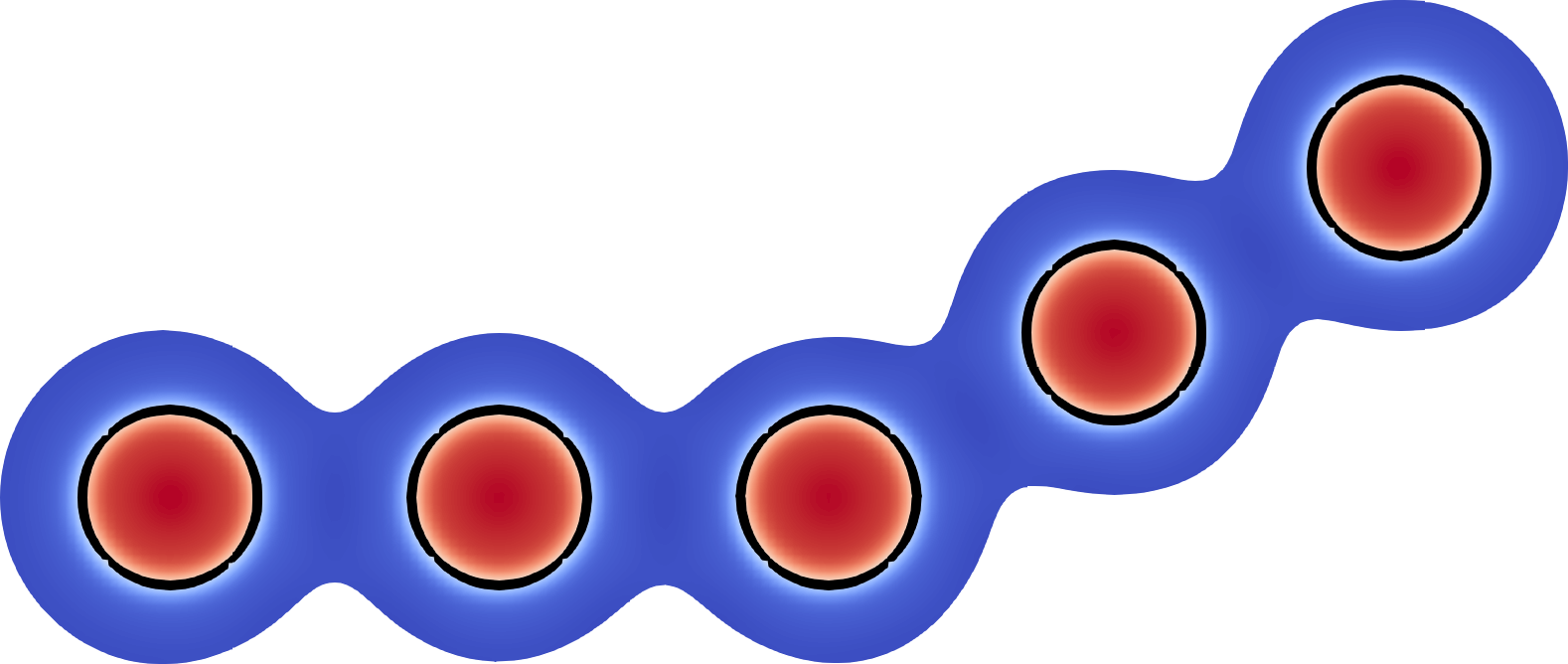} 
\\ \hline
$4.00$ &
\includegraphics[scale=0.07]{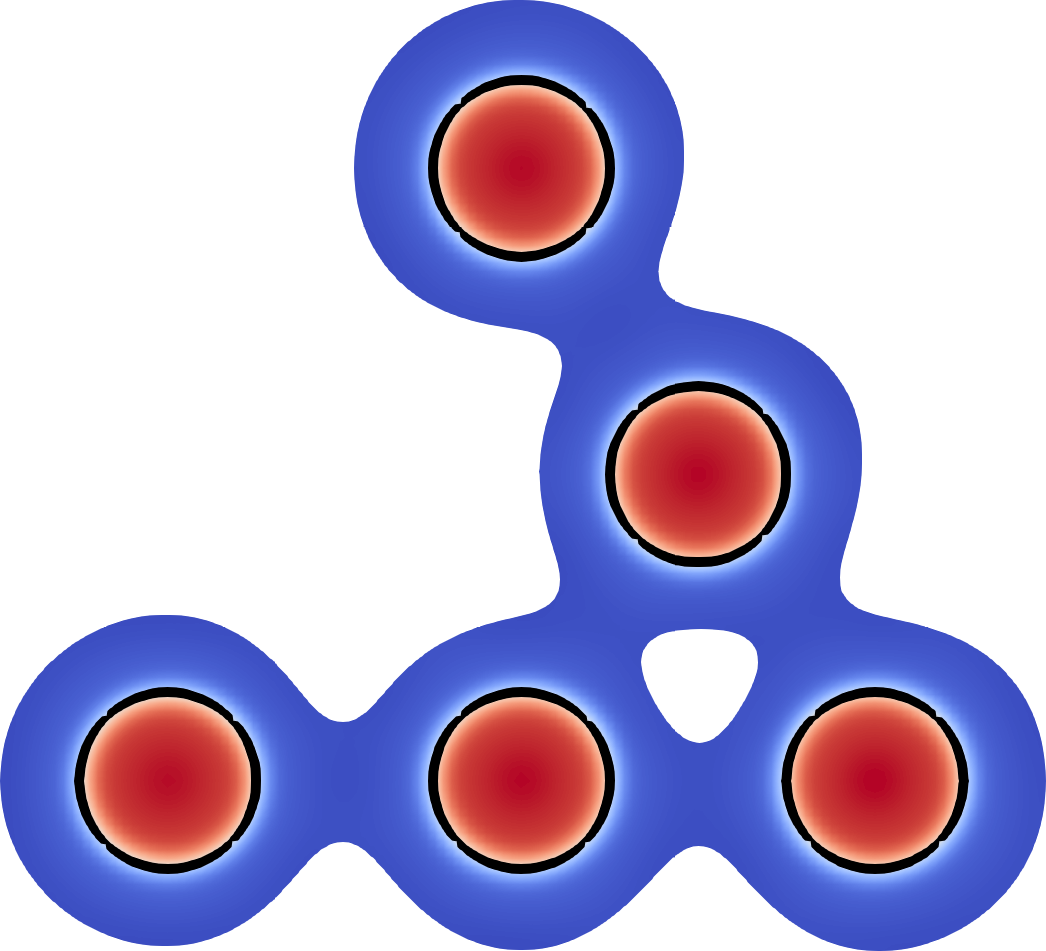} &
\includegraphics[scale=0.07]{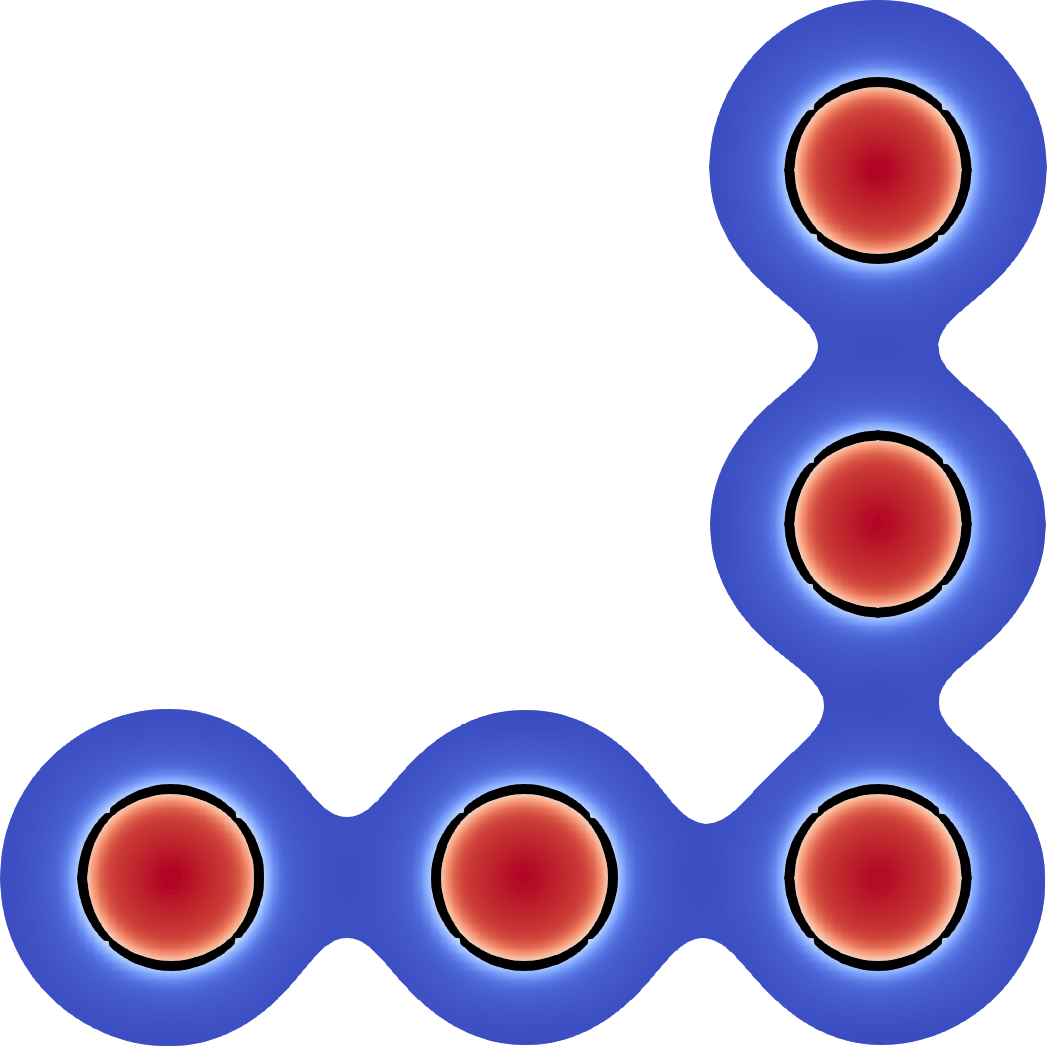} &
\includegraphics[scale=0.07]{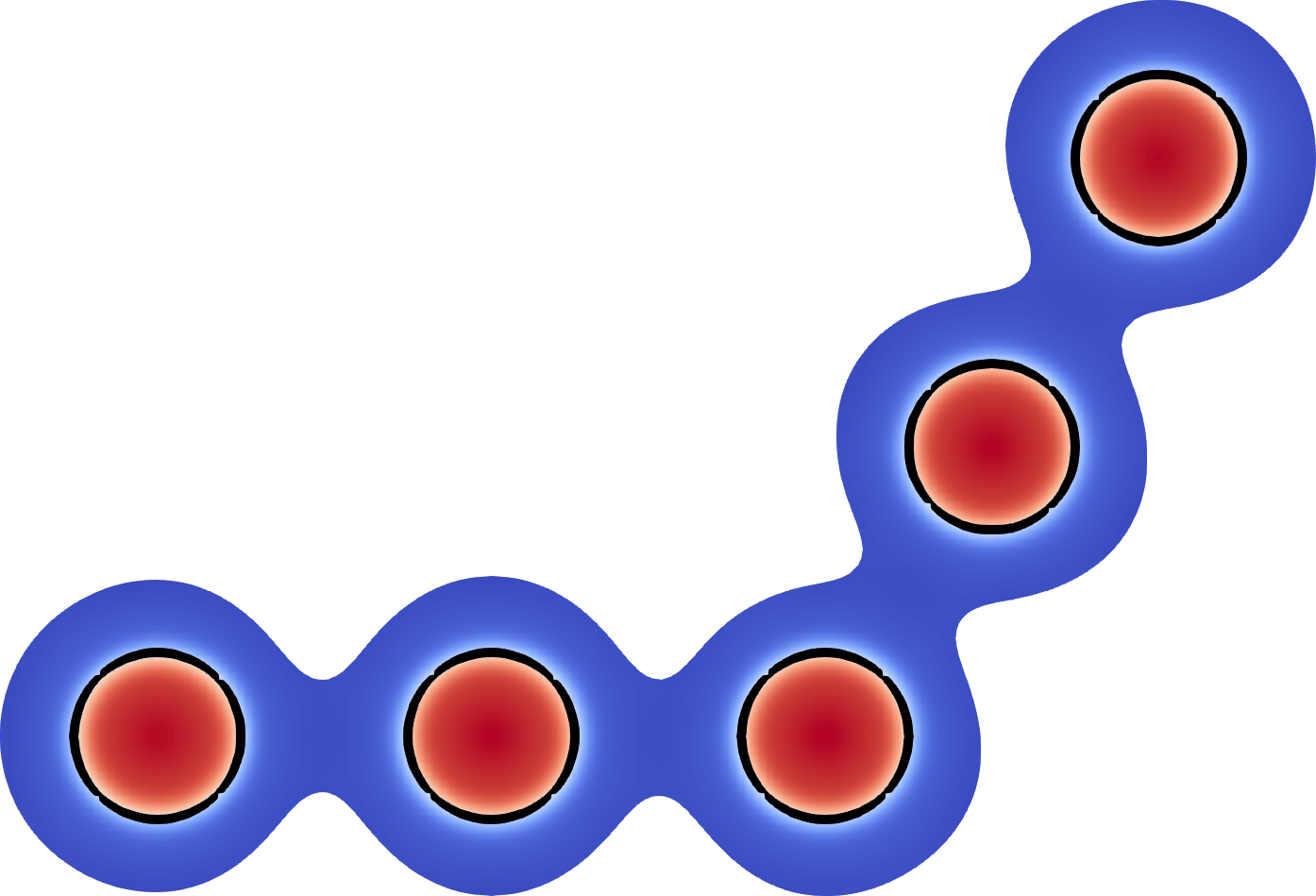} &
\includegraphics[scale=0.07]{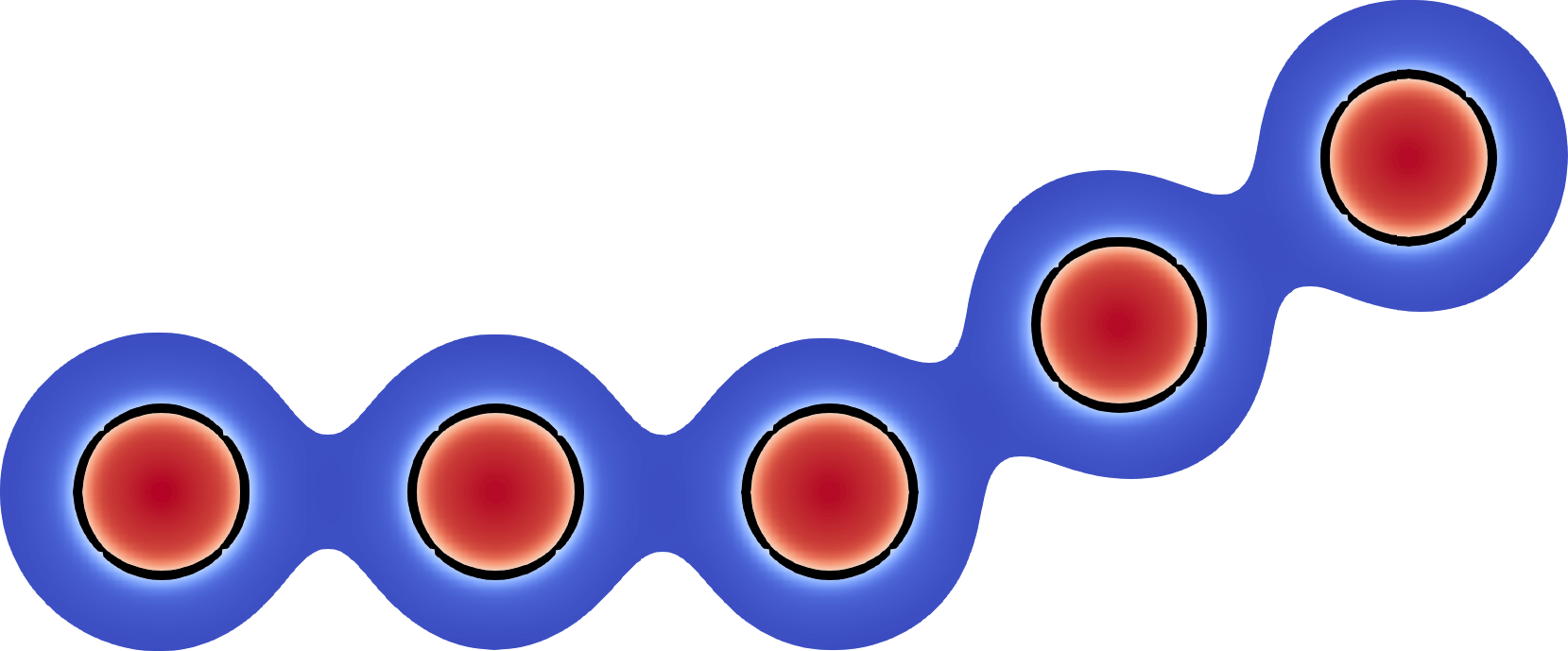} 
\\ \hline
\end{tabular}
\caption{Confining mask for the placement of a line of cylinders making a specified turn. Coloring shows the density configuration of the self-assembled polymer domains while the solid black line represent the target template.}
\label{fig:results:dsa:angle}
\end{figure}
As one can see, the proposed algorithm was able to find confining masks that result in very close matches between actual polymer morphologies and the desired ones, except for the case $\theta = 90^o$ and domain spacing $\Delta r = 3R_g$. However, we do not interpret this as a deficiency of the numerical method but as a more fundamental incommensurability between the desired template and the intrinsic properties of the polymer material considered. Indeed, in the cases of angles $\theta = 60^o$ and $\theta = 120^o$, which are more commensurate with the intrinsically favorable hexagonally packed polymer morphology, the inverse design is successful.
\subsection{Confining mask for more complex templates}
\begin{figure}[!h]
\centering
\begin{subfigure}[c]{\textwidth}
\centering
\includegraphics[scale=0.05]{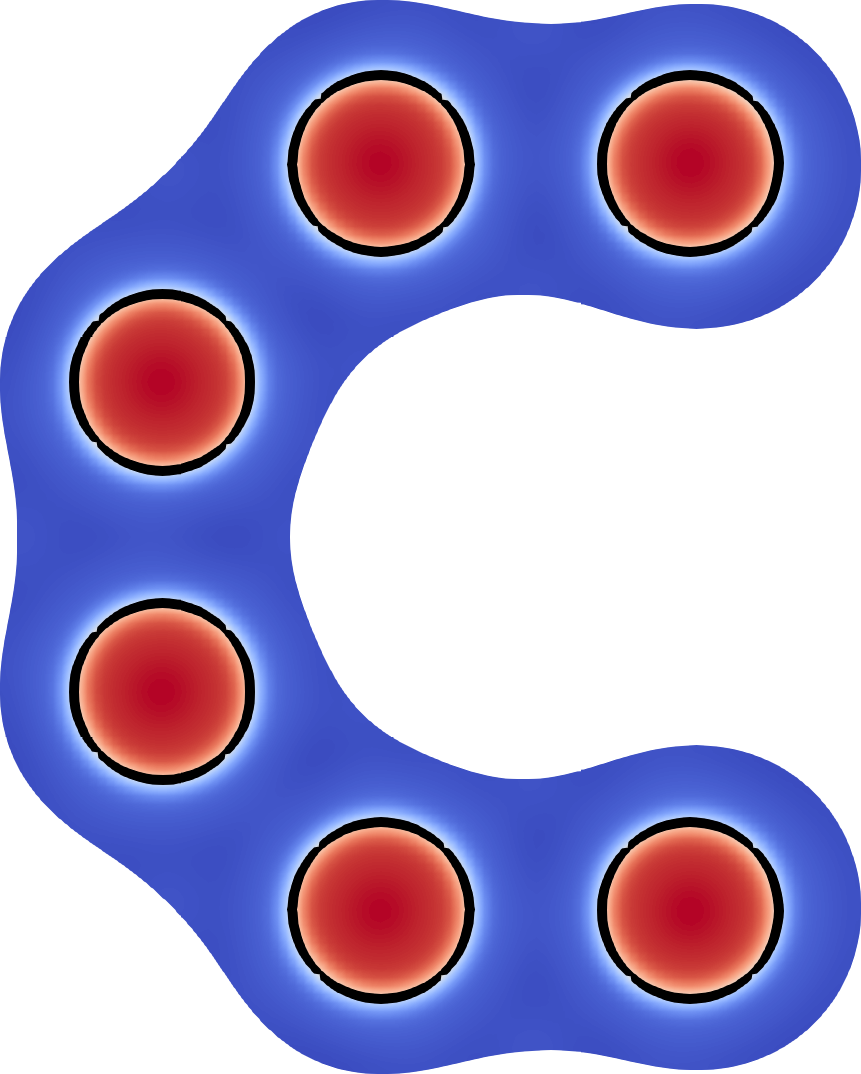} \,
\includegraphics[scale=0.05]{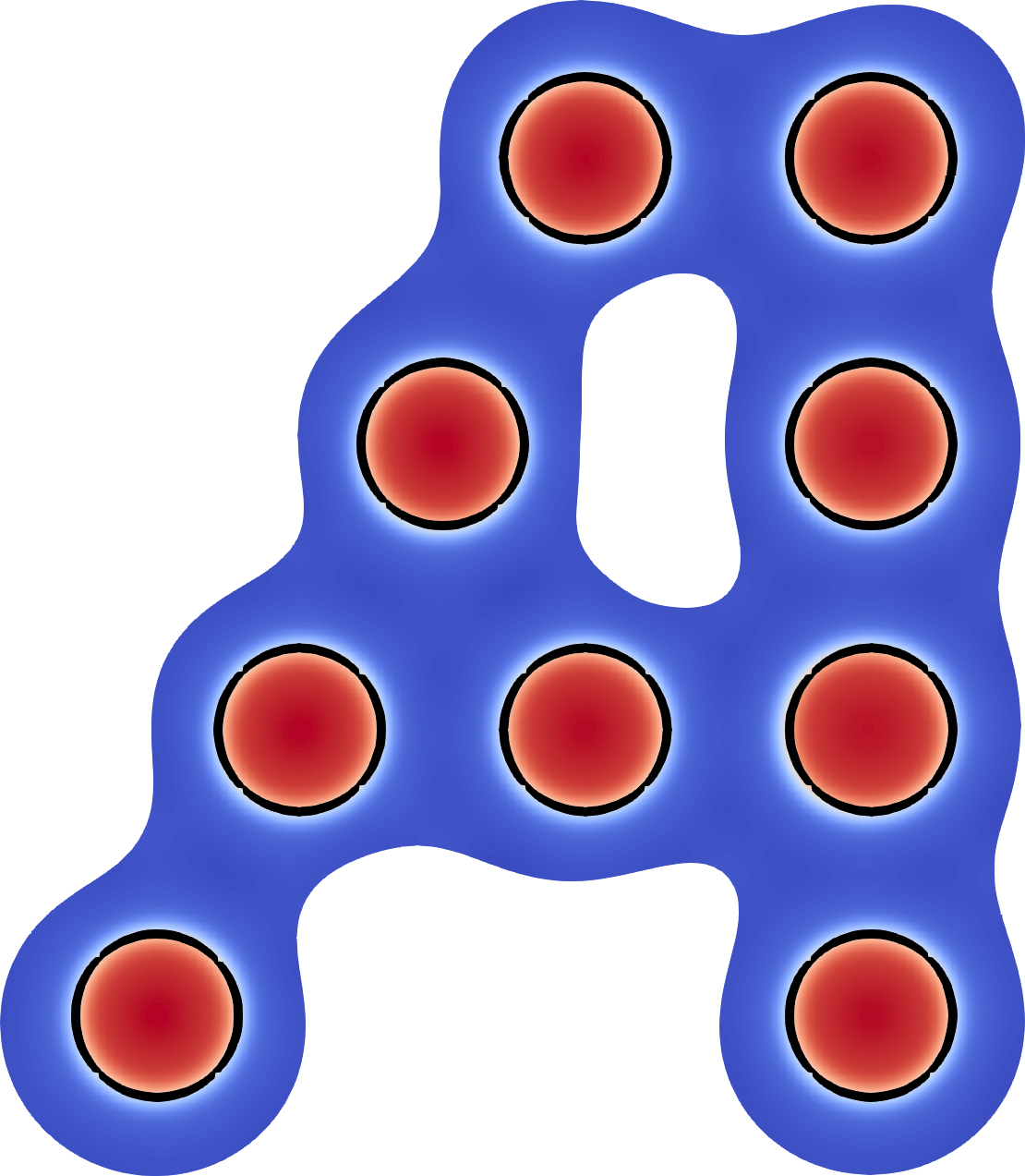} \,
\includegraphics[scale=0.05]{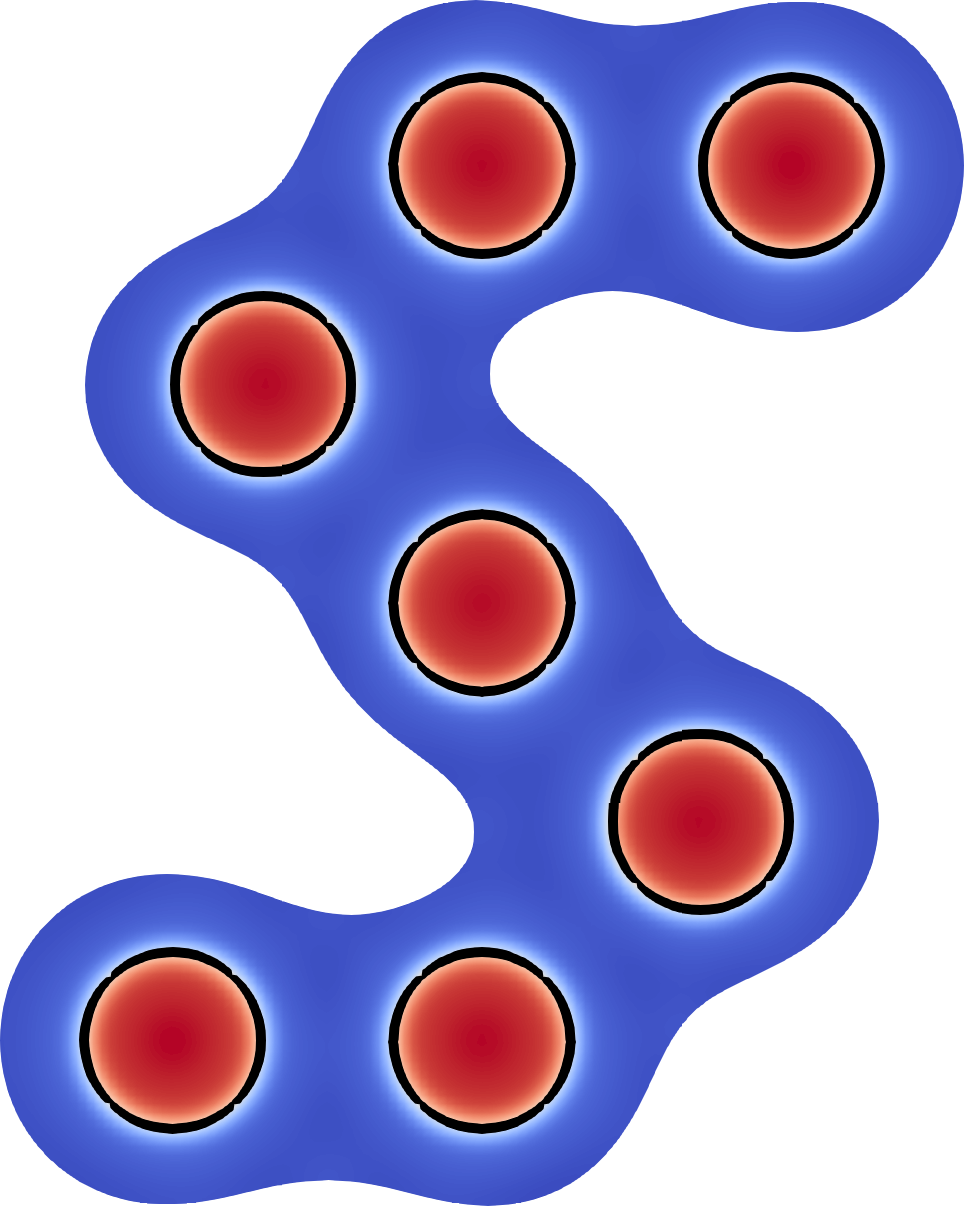} \,
\includegraphics[scale=0.05]{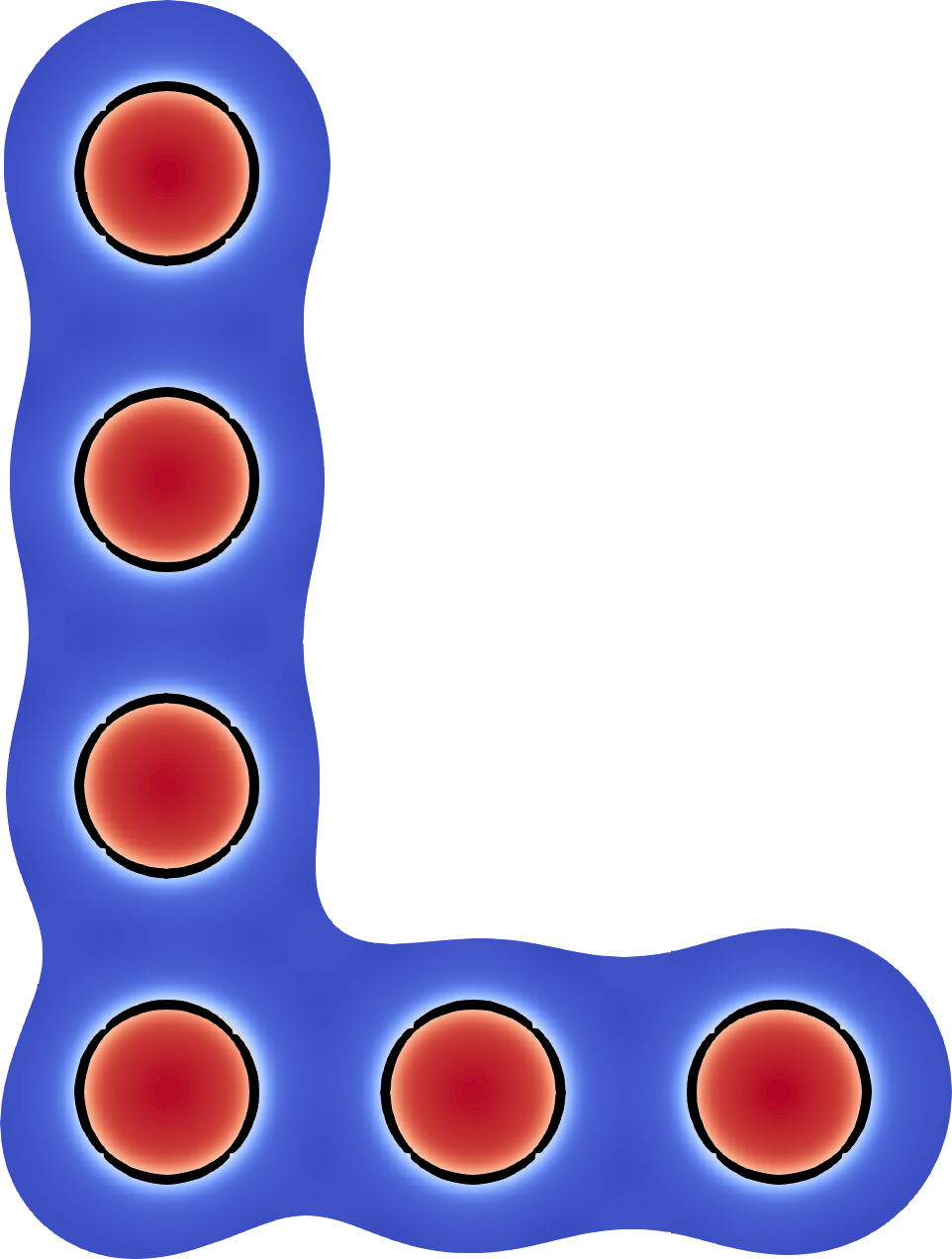} \,
\hspace{10pt}
\includegraphics[scale=0.05]{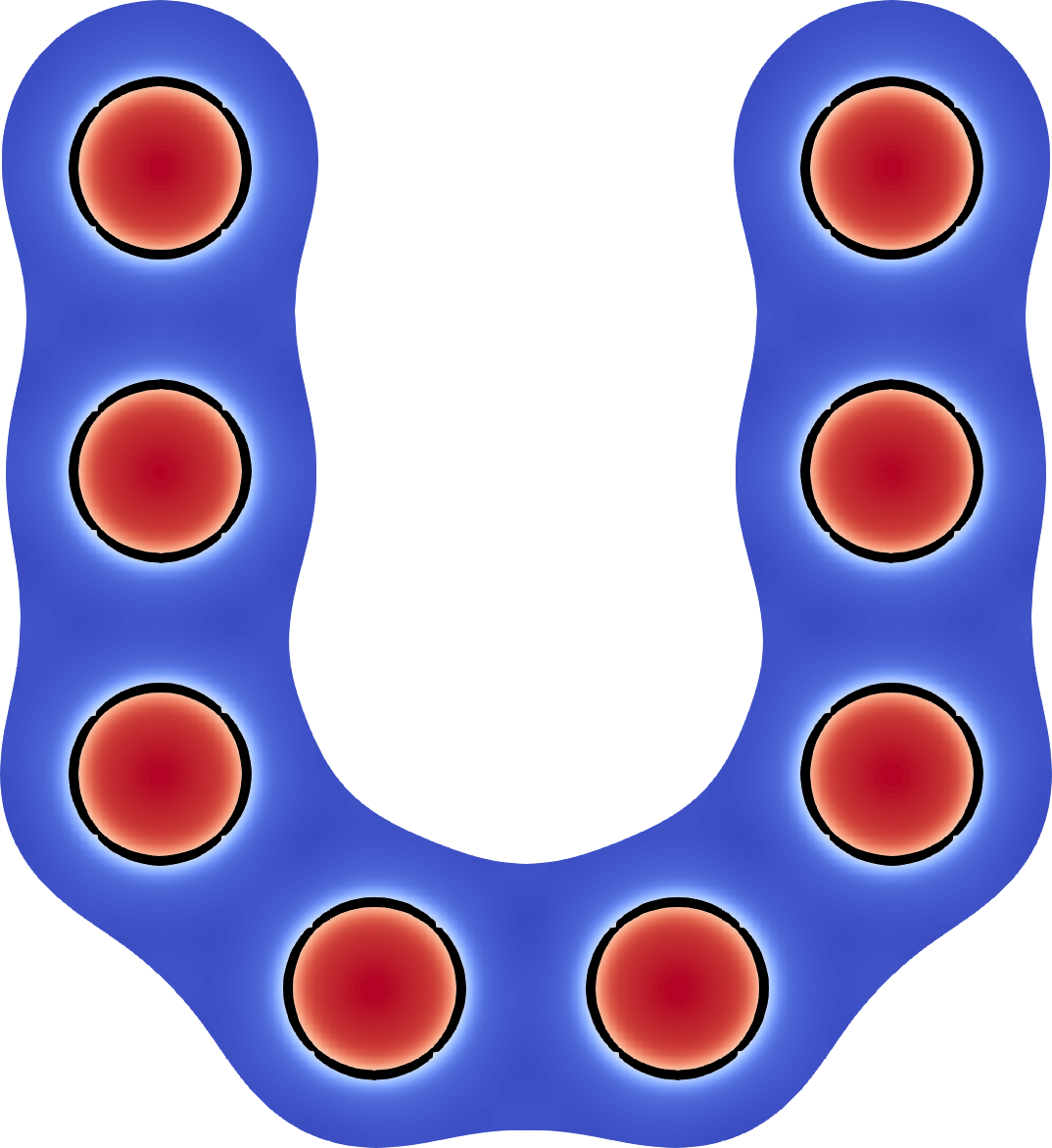} \,
\includegraphics[scale=0.05]{plot/dsa_examples/misc/dsa_c} \,
\includegraphics[scale=0.05]{plot/dsa_examples/misc/dsa_s} \,
\includegraphics[scale=0.05]{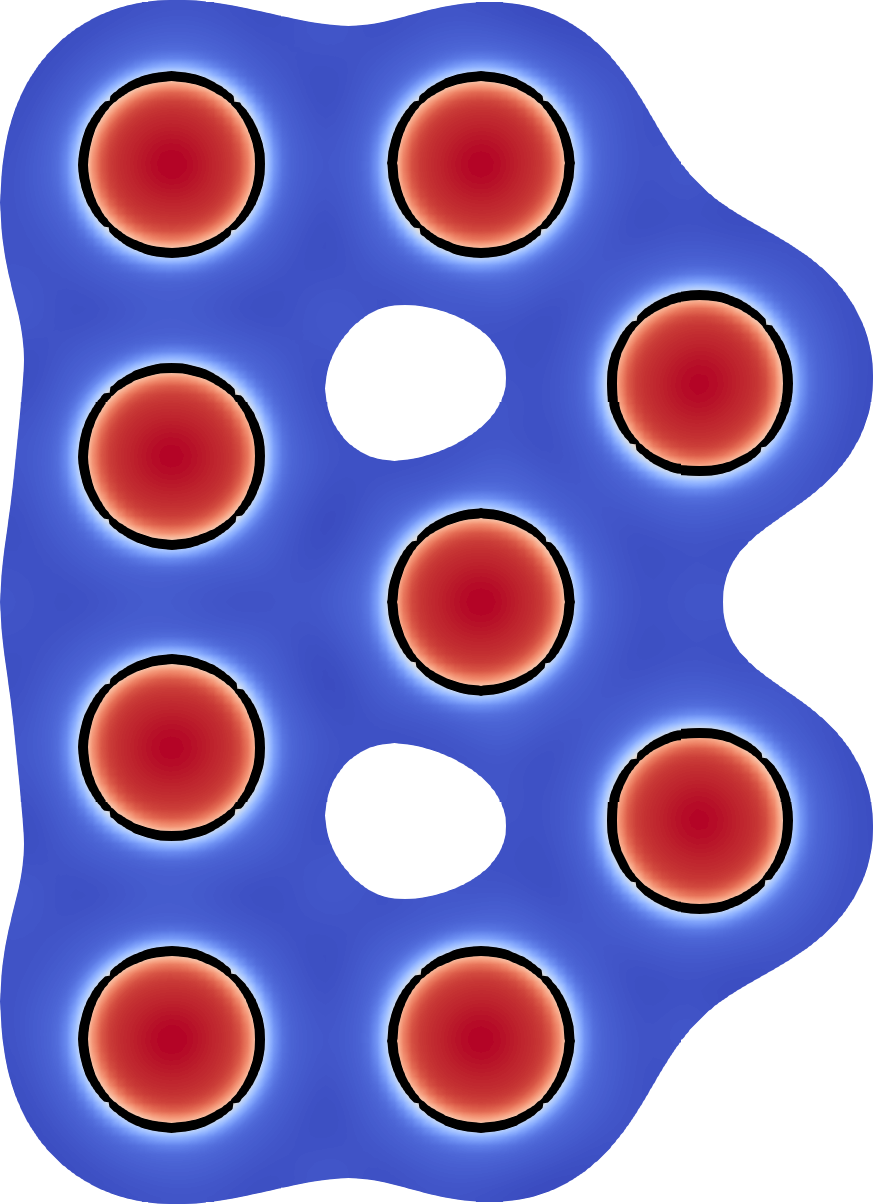}
\end{subfigure}
\caption{Confining masks for guiding cylindrical domains in the C, A, S, L, U, and B shaped patterns. Coloring shows the density configuration of the self-assembled polymer domains while the solid black line represent the target template.}
\label{fig:results:dsa:misc}
\end{figure}

Finally, in order to demonstrate the robustness and the flexibility of the proposed optimization algorithm, we apply it to the design of confining masks for more complex templates. Specifically, we consider the C, A, S, L, U, and B shaped patterns guiding 6, 9, 7, 6, 8, and 9 cylindrical domain, correspondingly.
In all cases, the characteristic distance between the cylindrical domains and their size are chosen to be $\Delta r = 3.5 R_g$ and $r_0 = R_g$. Figure \ref{fig:results:dsa:misc} illustrates the output of the optimization algorithm and demonstrates that successful and non-trivial confining masks are obtained for all patterns. 

\section{Conclusion}\label{sec:conclusion}

In this work we introduced an adjoint-based method for non-parametric shape optimization of guiding patterns in Directed Self-Assembly Lithography. The approach is based on using exact analytical sensitivities of the misfit between desired and actual morphologies of polymer self-assembly, which is modeled using Self-Consistent Field Theory. The method was applied to obtain confining masks for placement of cylindrical domains formed by a diblock copolymer in a number of different patterns. The method was shown to be able to produce masks that result not only in an accurate placement of cylindrical domains but accurate dimensions as well. Additionally, it was applied to cases of rather complicated patterns to demonstrate the method's robustness. Note, however, that the presented algorithm is aimed only at producing candidate geometries of guiding patterns that result in desired polymer morphologies and does not guarantee whether such morphologies are energetically preferred in these geometries. Improving the algorithm to take into account undesired competing polymer morphologies will be a topic of future research. Another direction of future work will be an integration with lithographic simulations predicting geometries of guiding patterns.

\subsection* {Acknowledgments}
The authors would like to acknowledge Gaddiel Ouaknin for useful discussions. This research was supported by NSF DMS 1620471.


\bibliography{references}   
\bibliographystyle{spiejour}   

\end{spacing}
\end{document}

%% file: packages.tex
\usepackage{amsmath, mathtools, environ}
\usepackage{amstext}
\usepackage{amssymb}
\usepackage{graphicx}
\usepackage{import}
\usepackage{appendix}
\usepackage{lscape}
\usepackage{subcaption}
\usepackage[boxed]{algorithm}
\usepackage{algorithm}
\usepackage{algpseudocode}
\usepackage{multirow}

\usepackage{longtable}

\usepackage{mleftright}

%% file: macros.tex
\newcommand{\pd}[2]{\frac{\partial #1}{\partial #2}}

\newcommand{\vect}[1]{\boldsymbol{#1}}

\newcommand{\hf}{{\frac12}}

\newcommand{\diff}[1]{\, d#1}

\usepackage[dvipsnames]{xcolor}

\newcommand{\mylinelabel}[1]{}
\newcommand{\abs}[1]{\left| #1 \right|}
\newcommand{\ddn}[2][]{\partial_{\vect{n}_{#1}} #2}

\newcommand{\lap}{\nabla^2}

\newcommand{\rin}[1][\vect{r}]{\vect{r} \in}

\newcommand{\vn}{v_{\vect{n}}}
\newcommand{\vecr}{\vect{r}}

\newcommand{\of}[1]{\mleft( #1 \mright)}

\newcommand{\fd}[2]{\frac{\delta #1}{\delta #2}}

\newcommand{\mum}{\mu_-}
\newcommand{\mup}{\mu_+}
\newcommand{\qf}{q}
\newcommand{\qb}{q^\dagger}

\newcommand{\deriv}[2]{\frac{d #1}{d #2}}
\newcommand{\ham}{\mathcal{H}}
\newcommand{\lag}{\mathcal{L}}
\newcommand{\Q}{\mathcal{Q}}
\newcommand{\lf}{\lambda}
\newcommand{\lb}{\lambda^\dagger}
\newcommand{\lm}{\lambda_-}
\newcommand{\lp}{\lambda_+}
\newcommand{\dds}[1]{\partial_s #1}
\newcommand{\myint}{\int\limits}

\newcommand{\cost}{\mathcal{F}}
\newcommand{\XN}[1]{\chi_{#1} N}
\newcommand{\sgn}{\textrm{sign}}

\newenvironment{myalign}%
{\par\begingroup\setstretch{1.0}\csname linenomath*\endcsname \align}%
{\endalign \csname endlinenomath*\endcsname \endgroup}

\newenvironment{myalign*}%
{\par\begingroup\setstretch{1.0}\csname linenomath*\endcsname \csname align*\endcsname}%
{\csname endalign*\endcsname \csname endlinenomath*\endcsname \endgroup}

\newenvironment{mymultline}%
{\par\begingroup\setstretch{1.0}\csname linenomath*\endcsname \multline}%
{\endmultline \csname endlinenomath*\endcsname \endgroup}

\newenvironment{mymultline*}%
{\par\begingroup\setstretch{1.0}\csname linenomath*\endcsname \csname multline*\endcsname}%
{\csname endmultline*\endcsname \csname endlinenomath*\endcsname \endgroup}

{\par\begingroup\setstretch{1.0}\csname linenomath*\endcsname \alignat{#1}}%
{\endalign \csname endlinenomath*\endcsname \endgroup}

\newenvironment{myalignat*}[1]%
{\par\begingroup\setstretch{1.0}\csname linenomath*\endcsname \csname alignat*\endcsname{#1}}%
{\csname endalignat*\endcsname \csname endlinenomath*\endcsname \endgroup}